\documentclass{article}

\usepackage{times,mathptm}
\usepackage{amsmath, amsfonts, bm}
\usepackage{graphicx}
\usepackage{natbib}
\usepackage{authblk}

\usepackage{fancyhdr}

\usepackage{algorithm}
\usepackage[noend]{algpseudocode}

\newlength{\npaperwidth}
\def\bfa{\mathbf{a}}
\def\bfb{\mathbf{b}}
\def\bfW{\mathbf{W}}

\usepackage{amssymb}
\usepackage{amsbsy}

\usepackage{noitcrul}
\newcommand\labeq[1] {\label{eq:#1}}

\newcommand\refeq[1] {\ref{eq:#1}}

\newcommand\eqp[1] {(\ref{eq:#1})}
\newcommand\eqeqp[1] {Equation~(\ref{eq:#1})}
\newcommand\eqseqp[1] {Equations~(\ref{eq:#1})}
\newcommand\Eqeqp[1] {Equation~(\ref{eq:#1})}

\renewcommand\,{\kern .075em}
\renewcommand\;{\kern .125em}

\newcommand\wrt{with respect to}
\newcommand\rhs{right-hand side}

\newcommand\pmf{probability mass function}
\newcommand\pdf{probability density function}
\newcommand\cdf{cumulative distribution function}
\newcommand\nn{neural network}
\newcommand\nns{neural networks}

\newcommand\NN{Neural Network}

\newcommand\E{\mathop\mathbb{E}\nolimits}

\def\calK{{\cal K}}
\def\calL{{\cal L}}

\def\bfx{\mathbf{x}}

\def\bfw{\mathbf{w}}

\def\bfgamma{{\boldsymbol{\gamma}}}
\def\bftheta{{\boldsymbol{\theta}}}

\def\pbar{{\bar p}}
\def\Pbar{{\bar P}}

\def\dee{\mathrm{d}}
\def\abd#1{|\dee #1|}
\def\pri{\mathsf{pri}}
\def\pos{\mathsf{pos}}
\def\worst{\mathsf{worst}}
\def\VB{\mathsf{VB}}
\def\ML{\mathsf{ML}}

\def\KL{\mathsf{KL}}
\def\DKL{D_\KL}

\newcommand\cnd {\,|\,}

\def\dot{\kern-.3ex\cdot\kern-.3ex}

\def\xp {^\mathit{T}}

\begin{document}

\title{\Large Bayesian Neural Networks for Geothermal Resource Assessment: \\
  Prediction with Uncertainty
}

\author[1,5]{Stephen Brown}
\author[1]{William L. Rodi}
\author[2,1]{Marco Seracini}
\author[1,6]{Chen Gu}
\author[1]{Michael Fehler}
\author[3]{James Faulds}
\author[3,7]{Connor M. Smith}
\author[4]{Sven Treitel}

\affil[1]{Department of Earth, Atmospheric, and Planetary Sciences \linebreak
  Massachusetts Institute of Technology, Cambridge, MA USA}

\affil[2]{Department of Physics and Astronomy ``Augusto Righi'', \linebreak
  University of Bologna, Bologna, Italy}

\affil[3]{Great Basin Center for Geothermal Energy, Nevada Bureau of
  Mines and Geology \linebreak University of Nevada, Reno, NV USA}

\affil[4]{TriDekon, Inc.,
  Tulsa, OK USA}

\maketitle

\footnotetext[5]{also at Aprovechar Lab L3C,
  Montpelier, VT USA}

\footnotetext[6]{now at Department of Civil Engineering
Tsinghua University, Beijing, China}

\footnotetext[7]{now at Zanskar Geothermal \& Minerals, Inc.,
  Salt Lake City, UT USA}

\begin{abstract}
  We consider the application of machine learning to the evaluation of
  geothermal resource potential. A supervised learning problem is
  defined where maps of 10 geological and geophysical features within
  the state of Nevada, USA are used to define geothermal potential
  across a broad region. We have available a relatively small set of
  positive training sites (known resources or active power plants) and
  negative training sites (known drill sites with unsuitable
  geothermal conditions) and use these to constrain and optimize
  artificial neural networks for this classification task. The main
  objective is to predict the geothermal resource potential at unknown
  sites within a large geographic area where the defining features are
  known. These predictions could be used to target promising areas for
  further detailed investigations. We describe the evolution of our
  work from defining a specific neural network architecture to
  training and optimization trials. Upon analysis we expose the
  inevitable problems of model variability and resulting prediction
  uncertainty. Finally, to address these problems we apply the concept
  of Bayesian neural networks, a heuristic approach to regularization
  in network training, and make use of the practical interpretation of
  the formal uncertainty measures they provide.
  
  \vspace*{\fill}
  %\centering -- Submitted to \em{Geophysics} --
  %\vskip 2\baselineskip
  %\centering Copyright \textcopyright\ 2023 Stephen R. Brown. All rights reserved.
  \centering Copyright \textcopyright\ 2023 Stephen R. Brown
  \vskip 3\baselineskip

\end{abstract}

\newpage
\tableofcontents

\newpage
\section{Introduction}

Many problems in Earth sciences concerning the environment, energy,
and natural hazards involve taking a set of observations on a map and
making predictions about some unknown or unseen characteristic: such as
predicting a numerical quantity, recognizing a particular structure or
category, or building a new classification or taxonomy. Applications
of practical importance are energy and mineral prospecting and
resource assessment, location of buried objects of specific types such
as land mines and tunnels, prediction of pathways for migration of
groundwater contaminants, or searching for anomalies within a complex
background as in security monitoring, to name a few.

To tackle these problems, one is tempted to bring to bear the recent
developments and successes in artificial intelligence (AI) and its
popular implementations known as machine learning, especially
artificial neural networks. However, the nature of the data in many
Earth science applications makes direct application of many successful
AI methods difficult.

First, our problems generally have a paucity of examples with known
labels (training sites) for problems for which we would hope to apply
supervised learning methods. Many problems in computer vision such as
facial recognition, autonomous driving, image recognition, etc., have
many tens to perhaps hundreds of thousands of labeled examples from
which to train, develop, and test deep network algorithms and
architectures. While some geoscience problems such as hyper-spectral
satellite imaging have as much training data
\citep[e.g.][]{Zhu:2017}, many have fewer than a hundred labeled
examples.  Another consideration is the nature of the data
itself. Input features carrying important information for the
regression or classification will necessarily be a mixture of
numerical values (real numbers such as temperature, distance from a
geologic fault, or gravity anomaly), categorical variables (mineral
assemblages or rock type descriptions), and ordinal variables (i.e.,
ranked as this is bigger than that, but with no scaling). Geologic
data will commonly be indexed to maps and thus are related spatially
to other features. These data, however, may not have the same
resolution nor the same degree of certainty.  Finally, there are
physical principles, variably understood ahead of time, which point to
reasonable relationships between features and labels. Incorporating
such ``expert'' knowledge into the hypotheses is important, especially
to counter the problem of a small number of examples and to ensure
proper weighting of the uncertainty of various data sources.

Generally, supervised machine learning uses two sources of knowledge:
(1) labeled data pairs (features, labels) and (2) hand-engineered
features, network architecture, and other components. For the cases of
relatively few labeled feature-outcome pairs, we often use more
hand engineering and/or specialized algorithms. For the cases having a
large number of training examples, we can generally apply simpler
algorithms and less hand engineering. For small numbers of training
data, hand engineering is commonly the most efficient way to improve
learning performance.

One would like to strike a balance between the ideal of feeding large
numbers of features into a network and letting the algorithm determine
the relationships and, on the other extreme, engineering the complete
hypothesis and algorithm by hand. The former, while unbiased and
allowing the data to speak for themselves, can be prone to overfitting
and for this application is perhaps at best inefficient or at worst
intractable. The latter runs the risk of extreme bias leading to
under-fitting such that important links among features will never be
recognized. Also, the algorithms developed in the hand-engineered
approach may not be extensible to new data types nor to new areas of
application. For these reasons, in this paper we use supervised
machine learning techniques and allow the training data examples to
determine the model details as much as possible.

Here, we consider the problem of prospecting and evaluation of
geothermal energy resources. We take as a case study the evaluation of
potential blind geothermal systems in the U.S. state of Nevada studied
previously in the Nevada Geothermal Play Fairway Project
\citep[][]{Faulds:2013, Faulds:2015, Faulds:2017, Faulds:2018,
  Faulds:2019} (see Figure \ref{fig_PFA_study_area}). We proceed by
revisiting the results from this previous study and re-analyzing the
accompanying data set using machine learning methods. In so doing, we
gain experience in bringing machine learning principles to bear on
this important class of Earth science problems. In the end we show
that by using a Bayesian approach we can naturally include model
uncertainty and provide a way not only to predict resource potential
but to also give measures of confidence or reliability.

\section{Problem Setting}

\begin{figure}[tb]
  \centering
  \includegraphics[width=0.8\textwidth]{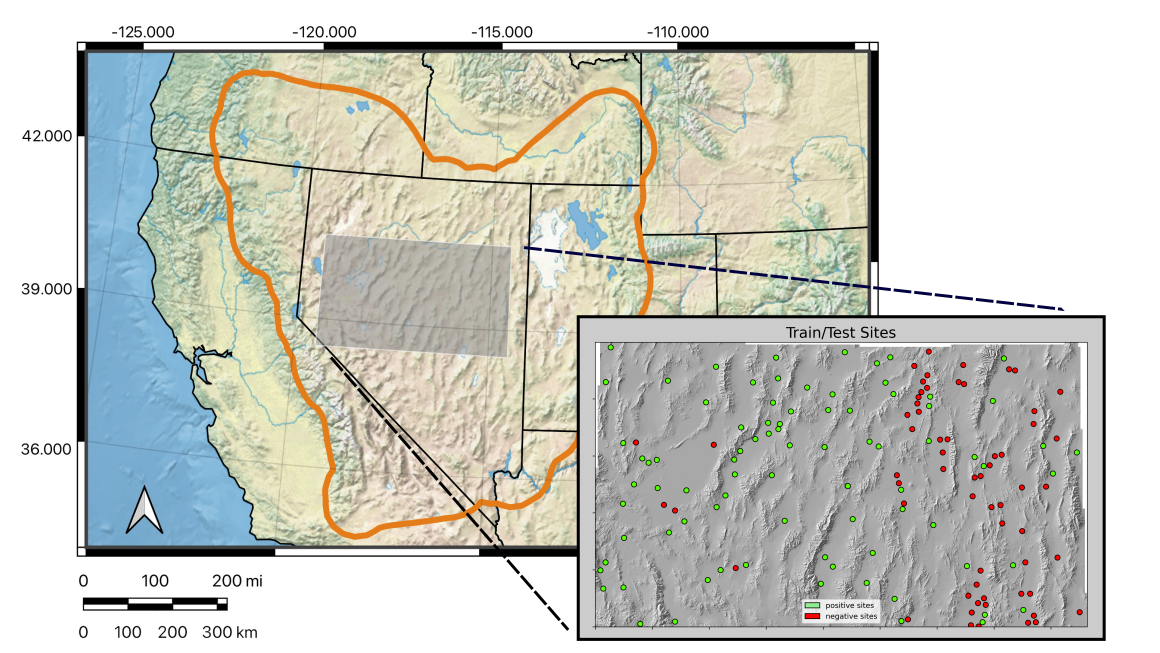}
  \caption{The Nevada Play Fairway Study Area (gray) is shown within
    the Great Basin Region (orange). The inset detail shows the
    locations of positive and negative training sites within the study
    area.} \label{fig_PFA_study_area}
\end{figure}

The Great Basin region (Figure
\ref{fig_PFA_study_area}) is a world-class geothermal province with
more than 1,200 MWe of installed capacity from 28 power
plants. Studies indicate far greater potential for both conventional
hydrothermal and enhanced geothermal systems (EGS) in the region
\citep[][]{Williams:2009}.

Because most geothermal systems in the Great Basin are controlled by
Quaternary normal faults, they generally reside near the margins of
actively subsiding basins. Thus, upwelling fluids along faults
commonly flow into permeable subsurface sediments in the basin and do
not reach the surface directly along the fault. Outflow from these
upwellings may emanate many kilometers away from the deeper source or
remain blind with no surface manifestations
\citep[][]{Richards:2002}. Blind systems are thought to comprise the
majority of geothermal resources in the region
\citep[][]{Coolbaugh:2007}. Thus, techniques are needed to both
identify the structural settings enhancing permeability and to
determine which areas may host subsurface hydrothermal fluid flow.

Geothermal play fairway analysis (PFA) is a concept adapted from the
petroleum industry \citep[][]{Doust:2010} that aims to improve the
efficiency and success rate of geothermal exploration and drilling by
integration of geologic, geophysical, and geochemical parameters
indicative of geothermal activity. A prior demonstration project
\citep[][]{Faulds:2017} focused on defining geothermal play fairways,
generating detailed geothermal potential maps for approximately 1/3 of
Nevada, and facilitating discovery of blind geothermal fields. The
Nevada Geothermal Play Fairway Project incorporated around 10
geologic, geophysical, and geochemical parameters indicative of
geothermal activity. It led to discovery of two new geothermal systems
\citep[][]{Craig:2018, Faulds:2018, Faulds:2019, Craig:2021}. The PFA
leveraged logistic regression, weights of evidence, and other
statistical measures as a type of machine learning technique. A set of
features, each gauged by a perceived weight of influence, were
combined to estimate geothermal potential. However, key limitations
and challenges affected the PFA, including estimating weights of
influence for parameters, limitations of some data sets, and a limited
number of training sites. We have been building upon the original
Nevada Geothermal Play Fairway Project of \citet[][]{Faulds:2017,
  Faulds:2018} to include principles and techniques of machine
learning \citep[][]{Faulds:2020, Brown:2020, Smith:2021}. In this
paper, we focus on the application of stochastic variational Bayes
neural networks to the problem of geothermal resource assessment.

\begin{figure}[tb]
  \centering
  \includegraphics[width=\textwidth]{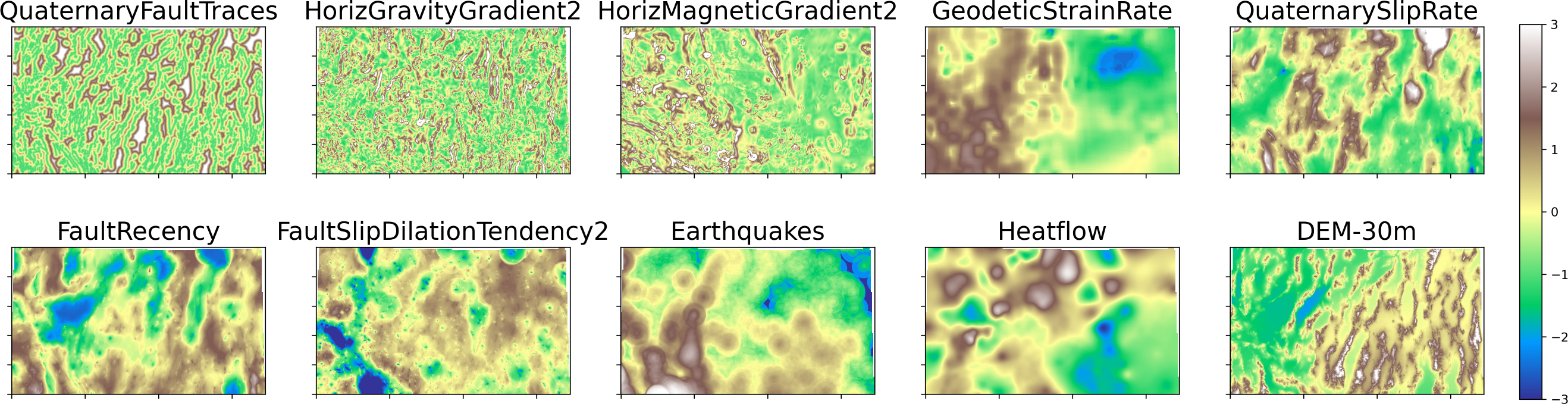}
  \caption{Geological and geophysical numerical feature maps used in
    this study. These maps were derived and updated from the original
    PFA study
    \citep[][]{Faulds:2018, Faulds:2020}.} \label{fig_PFA_numerical_features}
\end{figure}

\section{Classification with Fully Connected Neural Networks}

\subsection{Define a Supervised Learning Problem}

Our problem is constrained by data sets developed for the original
Nevada Geothermal Play Fairway Project. Some of these data were
further improved and made available for our use by
\citet[][]{DeAngelo:2021}.  Each of the geological and geophysical
features shown in Figure \ref{fig_PFA_numerical_features} are
presented as geo-referenced maps and are known everywhere on a
$250 \times 250$ meter pixel grid throughout the study area. At
certain geographic locations there are benchmarks, where it is known
whether or not a geothermal potential exists, i.e., an existing power
plant or definitive positive or negative drilling results. Training
data include 83 positive sites from known geothermal systems
($\geq 39^{\circ}$C) and 62 negative sites from wells that that showed
negative geothermal development potential, most of which were selected
from relatively deep and cool wells associated with oil and gas
exploration. The criteria used to select these training sites is
described by \citet[][]{Smith:2021}. Together these positive and
negative sites as shown in Figure \ref{fig_PFA_study_area} comprise
the “labels” used in training of the algorithms.

A supervised learning algorithm is trained by repeatedly presenting it
with these labeled examples and updating the model parameters until the 
predictions of the model match the training data satisfactorily. Ultimately
the trained model would then be used to predict the probability that
non-benchmark sites throughout the study region are positive geothermal
prospects.

In what follows we provide the mathematical foundation for
the supervised learning classification problem using artificial neural
networks (ANN). This motivates further descriptions of the construction and
training of the models, helps develop insight into the effects of
hyperparameter choices, and most importantly lends clear meaning to
the predictions.

%%%%%%%%%%%%%%%%%%%%%%%%%%%%%%%%%%%%%%%%%%%%%%%%%%%%%%%%%%
% BEGIN Bill's 1st Section
%%%%%%%%%%%%%%%%%%%%%%%%%%%%%%%%%%%%%%%%%%%%%%%%%%%%%%%%%%

\subsection{Mathematical Formulation for ANN}
\label{sec:mf_ann}

\subsubsection{Classification Problem}

We assume the existence of a population of potential
geothermal sites, each associated with a pair $(\bfx,y)$,
where the vector $\bfx$ contains numerical measurements
(features) acquired at the site, and where the Boolean
variable $y\in\{0,1\}$ indicates which of two classes the
site belongs to.  We take $y=1$ to mean the site is a
positive geothermal prospect and $y=0$ to mean it is a
negative prospect.  The geothermal classification problem is
to devise a function that maps the feature vector $\bfx$ to
the class $y$.

A statistical approach to this problem treats $y$ as a
random variable characterized by a \pmf\ (PMF), dependent
on $\bfx$.  We denote this PMF as $f(y\cnd\bfx)$,
showing the $\bfx$-dependence as conditioning even though
our approach will not consider the randomness of $\bfx$.
The PMF of $y$ takes values only for $y=0$ and $y=1$,
with $f(0\cnd\bfx)$ being the probability that $y=0$ for a
site with feature vector $\bfx$, and with
$f(1\cnd\bfx)=1-f(0\cnd\bfx)$.

Adopting the point of view of \citet[][]{Bishop:2006} and others, we
can break the classification problem into two subproblems.
The first, which is the topic of this paper, is to infer, or
learn, $f(y\cnd\bfx)$ from training data.  Given the
inferred $f(y\cnd\bfx)$, the second subproblem is to decide
whether $y=0$ or $y=1$ for a new site of interest having
feature vector $\bfx$.  A simple decision rule would be to
classify the site as $y=1$ if $f(1\cnd\bfx)>f(0\cnd\bfx)$
and as $y=0$ otherwise.  More complex rules ensue from
consideration of issues such as misclassification costs,
opportunity loss, and availability of prior information
about $y$.  These issues fall within the domain of decision
theory and are not addressed in this paper.

\subsubsection{Statistical Model}

We attribute the stochastic nature of $y$ to the notion of
random selection, whereby a site we wish to classify is
considered a random draw from a population of similar sites,
a proportion $p$ of which is labeled $y=1$ and the remainder
labeled $y=0$.  For a given value of $p$, the PMF of $y$
for the selected site is the Bernoulli distribution, given
by
\begin{align}
f(y\cnd p) &= p^y\,(1-p)^{1-y}  \labeq{bern}
\end{align}
or, enumerating,
\begin{align}
f(y\cnd p) &=
\begin{cases}
p & \text{for } y=1 \\
1-p & \text{for } y=0.  \labeq{bern2}
\end{cases}
\end{align}
The use of conditioning notation to indicate dependence on
$p$ is not gratuitous in this case, as we will later treat
$p$ as a random variable.  Implicit in the Bernoulli model
is that the dependence of $y$ on $\bfx$ arises solely from
the dependence of $p$ on $\bfx$, expressed as
\begin{align}
f(y\cnd p,\bfx) &= f(y\cnd p).  \labeq{phidesx}
\end{align}

Of importance to applications in decision theory we remark that, given
\eqp{bern2}, it is appropriate to refer to the parameter $p$ as a
probability, rather than a proportion of the population.  Further, in
the interpretation that the case $y=1$ is considered a ``successful''
draw, it is appropriate to consider $p$ to be the probability of
success.

The PMF $f(y\cnd\bfx)$ is determined by the Bernoulli
model in conjunction with a model for the dependence of $p$
on $\bfx$.  The latter model must be inferred from training
data and, since a training data set is inherently limited,
prior constraints on the relationship between $p$ and
$\bfx$.  Here, we constrain this relationship by taking $p$
to be a specified function of $\bfx$ and a vector of
parameters, $\bfgamma$, assumed to be independent of $\bfx$:
\begin{align}
p &= P(\bfx,\bfgamma).  \labeq{Pdef}
\end{align}
We construct the function $P$ as follows.  First, to enforce
$0\le p\le 1$, we let $p$ be the logistic function of an
unconstrained variable $z$:
\begin{align}
p &= \frac 1 { 1 + \exp\big(-z) }.  \labeq{logit}
\end{align}
We see that $z=-\infty$, $0$, $\infty$ map to $p=0$, 1/2, 1,
respectively.  Second, we let $z$ be a specified function of
$\bfx$ and $\bfgamma$, which we express as
\begin{align}
z &= Z(\bfx,\bfgamma).  \labeq{Zdef}
\end{align}
The method of logistic regression \citep[e.g.][Chapter
1]{Goodfellow:2016}, for example, takes $\bfgamma=(b,\bfw)$ and $Z$
as the linear function
\begin{align}
Z(\bfx,\bfgamma) &= b + \bfw\xp\bfx  \labeq{Zlogreg}
\end{align}
with $b$ a bias parameter and $\bfw$ a vector of
feature weights.  In general, combining \eqseqp{logit} and
\eqp{Zdef} defines the function $P$ in \eqp{Pdef}.

Our approach generalizes logistic regression by taking $Z$
to be a nonlinear function of $\bfx$, implemented as a
multilayer neural network with an input node for each
observed feature (component of $\bfx$) and a single output
node corresponding to $z$.  The components of $\bfgamma$ are
the biases assigned to the nodes in each layer and the
weights assigned to connections between the nodes of
adjacent layers.  Since the logistic function in \eqp{logit}
conforms to the activation functions typically used in
neural networks, it is convenient to, and we hereinafter do,
take the function $P$ itself to be the neural network, with
\eqp{logit} applied at the output node to yield $p$.

Combining \eqseqp{bern} and \eqp{Pdef}, we obtain an
expression for the PMF of $y$, now dependent on
$\bfgamma$ as well as $\bfx$:
\begin{align}
f(y\cnd\bfx,\bfgamma) &= P(\bfx,\bfgamma)^y \,
   \big( 1 - P(\bfx,\bfgamma) \big)^{1-y}.
  \labeq{pmfybarxgam}
\end{align}
The inference of $f(y\cnd\bfx)$ reduces to the inference of
the parameter vector $\bfgamma$.

\subsubsection{Training the Neural Network}

For a fixed value of $\bfgamma$, the function $P$ represents
a conventional, deterministic \nn\ that predicts, for
arbitrary $\bfx$, a specific value of $p$ (via \eqeqp{Pdef})
as well as the PMF of $y$ (via \eqp{pmfybarxgam}).  A
suitable value of $\bfgamma$ to use for these predictions
results from applying machine learning to a set of training
data obtained from sites with known class.  Given $N$ such
sites, we denote the training data set $(\bfx_i,y_i)$, $i=1$,
\dots, $N$, or, for brevity, the pair of tuples $(X,Y)$
defined as
\begin{align}
X &= (\bfx_1,\bfx_2,\ldots,\bfx_N)  \labeq{Xtuple} \\
Y &= (y_1,y_2,\ldots,y_N).  \labeq{Ytuple}
\end{align}
A standard approach in machine learning is to set $\bfgamma$
to the value that maximizes the likelihood of the training
data set, defined as $f(Y\cnd X,\bfgamma)$ and, considered as
a function of $\bfgamma$, known as the likelihood function.
With $Y$ being known, this is a supervised learning approach
and, given our explicit model parameterization, reduces to
statistical regression.

We assume the $y_i$ are statistically independent, implying
in conjunction with \eqeqp{pmfybarxgam}
\begin{align}
f(Y\cnd X,\bfgamma) &= \prod_{i=1}^N
P(\bfx_i,\bfgamma)^{y_i} \,
\big(1-P(\bfx_i,\bfgamma)\big)^{1-y_i}.  \labeq{like}
\end{align}
In numerical applications it is convenient to deal with the
{\em negative} logarithm of likelihood, which we show as a
function, $\calL$, of $\bfgamma$ and the training data.
\Eqeqp{like} yields this as
\begin{align}
\calL(\bfgamma;X,Y)
&\equiv - \log f(Y\cnd X,\bfgamma)
\nonumber \\
&= - \sum_{i=1}^N y_i\,\log P(\bfx_i,\bfgamma) -
   \sum_{i=1}^N (1-y_i)\,\log\big(1-P(\bfx_i,\bfgamma)\big)
\nonumber \\
&= - \sum_{i=1\atop y_i=1}^N \log P(\bfx_i,\bfgamma) -
   \sum_{i=1\atop y_i=0}^N \log\big(1-P(\bfx_i,\bfgamma)\big)
  \labeq{calL}
\end{align}
where the summations in the latter expression separate the
positive training sites ($y_i$=1) from the negative ones.
The maximum-likelihood value of $\bfgamma$ can then be
stated as
\begin{align}
\bfgamma_\ML &= \arg\min_\bfgamma \calL(\bfgamma;X,Y).  \labeq{gammaML}
\end{align}

Our implementation of deterministic \nns\ calculates
$\bfgamma_\ML$ by applying a gradient descent procedure to
minimize $\calL$, using the well-known back-propagation
technique \citep[e.g.][Chapter 6]{Goodfellow:2016} to perform gradient
computations.

\subsubsection{Regularization of Network Training}
\label{sec:ann_regularization}

\Eqeqp{gammaML} defines $\bfgamma_\ML$ as the value of
$\bfgamma$ that best predicts, or fits, the training
data set.  This means that, overall, $P(\bfx_i,\bfgamma_\ML)$
is as close as possible to one for positive training sites
and to zero for negative training sites.  Being optimal in
this sense, however, does not guarantee that $\bfgamma_\ML$
will successfully predict the class of new sites.  The
predictive value of $\bfgamma_\ML$ will ultimately depend on
two factors: how well the parameter vector $\bfgamma$
characterizes the relationship between $p$ and $\bfx$, for
relevant values of $\bfx$; and how well the training data set
constrains $\bfgamma$.  When $\bfgamma$ is not well
constrained $\bfgamma_\ML$ will be vulnerable to
``overfitting'' the training data and will yield poor
predictions for new values of the feature vector $\bfx$.

The classical remedy for overfitting in machine learning,
and inverse problems in general, is regularization.  In our
context, this means relaxing the maximum-likelihood
criterion in \eqeqp{gammaML} with auxiliary conditions on
$\bfgamma$.  A standard technique is to augment $\calL$ with
a penalty term and seek the value of $\bfgamma$ that
minimizes the augmented function.

%%%%%%%%%%%%%%%%%%%%%%%%%%%%%%%%%%%%%%%%%%%%%%%%%%%%%%%%%%
% END Bill's 1st Section
%%%%%%%%%%%%%%%%%%%%%%%%%%%%%%%%%%%%%%%%%%%%%%%%%%%%%%%%%%

\subsection{Neural Network Architecture}

The original Nevada Geothermal Play Fairway Project as applied to the
Great Basin devised a computation graph grounded in statistical
analyses for estimation of geothermal resource potential in the region
\citep[refer to Figure 3 of][]{Faulds:2018}. Specific geological and
geophysical features were chosen as they were thought to indicate
essential aspects of permeability and heat sources. These features
were combined algebraically in a method conditioned by the knowledge
of known resource benchmark sites to calculate the ``fairway,'' a
number scaled to indicate a region-wide degree of the geothermal
potential \citep[Figure 1 of][]{Faulds:2018}.

We recognized that this computation graph is a feed-forward network,
in effect a highly engineered neural network where the nodes are
connected in a very specific way.  In this paper we build upon this
previous work and formulate the problem using standard artificial
neural network techniques. We want to build upon the experience gained
in the previous project of \citet[][]{Faulds:2018}, therefore we use
the same feature data sets with enhancements \citep[][]{DeAngelo:2021}
allowing us to use their fairway map as a point of comparison with our
results.

\begin{figure}[h]
  \centering
  \includegraphics[width=0.4\textwidth]{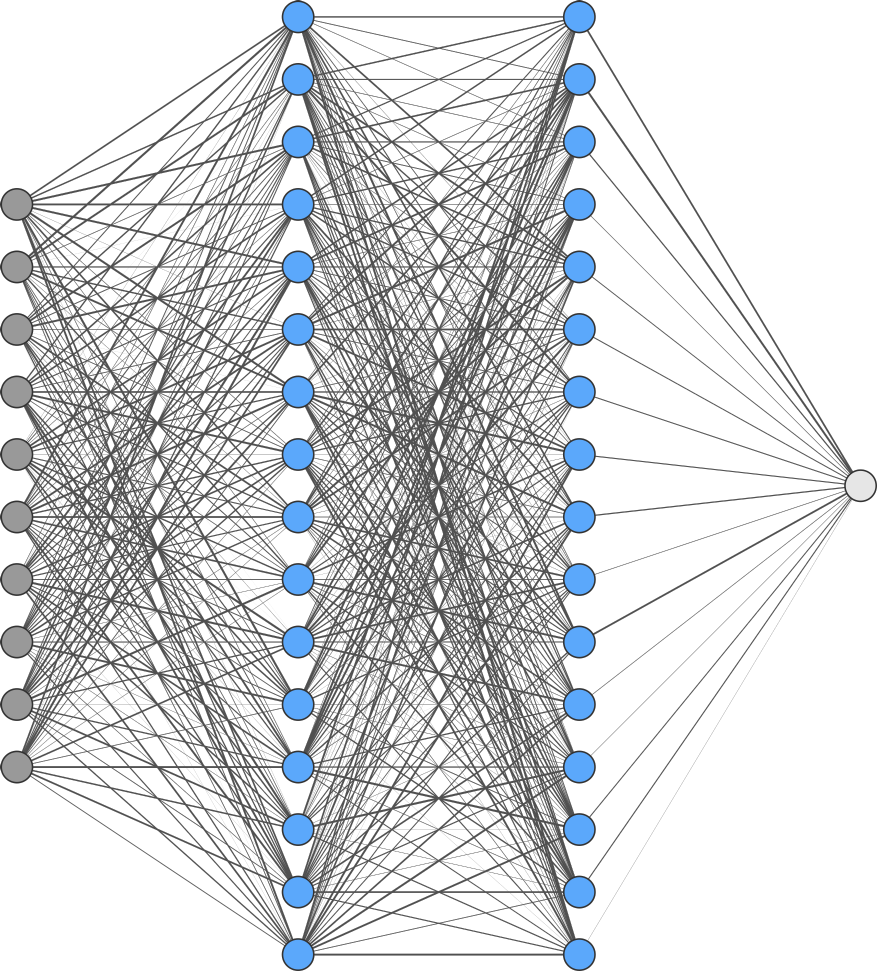}
  \caption{Fully connected neural network used for this study. The
    vertical set of dark gray circles on the left represents the 10
    input features, the light gray circle on the right represents the
    output neuron (prediction) and the vertically-aligned blue circles
    in the center represent the 16 neurons in each of the two hidden
    layers. Each shaded gray line indicates that each interconnection
    (combination of weights and biases) is unique.} \label{fig_FCNN}
\end{figure}

We first must decide on the specific fully connected network
architecture to use. In particular we must decide upon the number of
neurons per layer and the number of layers and choose values of any
additional hyperparameters the algorithm requires. A scoping study was
done to find an optimum set of model parameters. A ``best'' network
was sought using genetic algorithm methods (specifically the Python
implementation ``DEAP'' of \citet[][]{DEAP:2012}) to efficiently
search through various network structures and combinations of
hyperparameters. The suitability of each candidate was gauged through
performance metrics that emphasize the trade-off between data
over-fitting and model complexity (the Akaike Information Criterion
and the Bayesian Information Criterion described in
\citet[][]{AIC_BIC:2014}). The genetic algorithm built a population of
networks, evaluated each one, modified them, and kept the most
suitable individuals. In the end we found that there are certainly
unsuitable networks to use. Those networks too simple, such as one
with only one hidden layer only or a network mimicking the engineered
computation graph used in the original Nevada Geothermal Play Fairway
Project \citep[Figure 3 of][]{Faulds:2018} or those networks with very
narrow yet deep configurations (such as 8 layers of 4 neurons each,
for example) do not train consistently nor generalize well to new
examples not seen during training. At the other extreme, those
networks both wide and deep (such as 8 layers of 16 neurons each, for
example) are also unsuitable since these have large numbers of
parameters, overfit or memorize the training data, and must be heavily
regularized in order to generalize. There are, however, many networks in
between which seem to be equally suitable as they train consistently
well, can fit our the data to reasonable accuracy and can be easily
kept from overfitting with some form of regularization. The network we
settled upon for our studies is shown in Figure \ref{fig_FCNN} with
two hidden layers of 16 neurons each. The number of neurons was chosen
to be just more than the anticipated maximum number of features to be
available for study, thinking that it would allow all conceivable
interconnections among the features in determining the outcome. The
single output neuron indicates the probability of a positive site.

\subsection{Implementation}
\label{sec:implement_section_ann}

We provide a brief description of the implementation of a basic
fully connected neural network. As we follow standard methods in the
field of machine learning, we will only highlight those algorithm
components helpful for a comparison of this network to the
probabilistic networks employed later.

We read from left to right in the graphical representation of the
network in Figure \ref{fig_FCNN}. The network consists of a set of
neurons organized into layers. The set of input features are combined
linearly at each neuron and then passed through a nonlinear activation
function. The output of each activation function becomes a new internal
feature for the next layer of neurons. This computation is applied,
layer upon layer, until the final neurons are reached. The outputs of
the final layer represent the prediction. Such a forward pass through the
network is described in pseudo code as Algorithm
\ref{forward_pass}. The forward-pass algorithm is at the heart of both
training of the network and its subsequent use for prediction.

\centerline{%
\begin{minipage}{0.9\textwidth}
\begin{algorithm}[H]
%\begin{algorithm}
  \caption{Forward pass through a neural network computing
    $p = P(\bfx,\bfgamma)$, with $\bfgamma = (\bfW,\bfb)$ from
    Equation \eqp{Pdef}. Inputs are given as a column vector $\bfx$
    containing all of the input features. For each neuron we also
    define a column vector of weights $\bfw$, each component of which
    corresponds to a specific input to that neuron. Expanding from the
    simple logistic regression case to layers of many neurons, we now
    represent each layer by the weight matrix $\bfW$ containing the
    horizontally stacked weight vectors $\bfw$ for all neurons in that
    layer. Once transposed, each row of $\bfW^{T}$ represents a neuron
    and the elements of each column represent the weights multiplying
    the output of each neuron in the previous layer. A column vector
    $\bfb$ contains the biases for all neurons. Each step represents a
    set of linear combinations of the outputs from the previous layer
    ($\bm{z}$) represented as a matrix algebra operation followed by
    the application of a nonlinear activation function $g(\bm{z})$,
    possibly unique to each layer. The network returns the predictions
    as the activations from the final layer.}
  \label{forward_pass}
  \begin{algorithmic}[1] % The number tells where the line numbering should start
      \Procedure {ForwardPass}{$\bfx, \bfW, \bfb$}
        \State $\bfa^{[0]} \gets \bfx$ \Comment{begin with the input features}
        \For{$l$ in layers} \Comment{loop over all layers}        
          \State $\bm{z}^{[l]} \gets \bfW^{[l] \textrm{T}} \bfa^{[l-1]} + \bfb^{[l]}$
            \Comment{linear combinations of previous layer's outputs}
          \State $\bfa^{[l]} \gets g^{[l]} ( \bm{z}^{[l]} )$ \Comment{apply activation function}
          \EndFor
          \State $p \gets \bfa^{[\textrm{final}]}$ \Comment{final layer's activation is the probability}
        \State \textbf{return} $p$ \Comment{return the prediction}
      \EndProcedure
  \end{algorithmic}
\end{algorithm}
\end{minipage}
\vspace{2\baselineskip}
}

The main steps in training a neural network are summarized in pseudo
code in Algorithm \ref{training_fcnn}. Briefly, to train the network
first any model hyperparameters such as regularization parameters are
chosen. Then the trainable model parameters (the weights and biases)
are initialized to random values. A set of training examples is
repeatedly presented to the algorithm. The training features are used
in Algorithm \ref{forward_pass} to produce an interim prediction. This
prediction is then compared to the known outcome for the example and
the misfit between them, known as the loss or the cost, is used in an
optimizer for and incremental update of the model parameters. This
training loop proceeds until a satisfactory solution is found and the
final model parameters are saved. The training process is monitored
through learning curves consisting of both the evolution of the loss
and the evolution of the predictive accuracy. These training metrics
are reported each epoch. An epoch counts one forward pass of every
training example through the network. Some training examples are held
out and used to test the ability of the network to generalize to
unseen examples.

\centerline{%
\begin{minipage}{0.9\textwidth}
\begin{algorithm}[H]
  \caption{Training a neural network. A set of example features
    $\bm{X}$ and corresponding labels $\bm{Y}$ are used to iteratively
    improve the model weights $\bfw$ and biases $\bfb$ from their
    initial random values by comparing model predictions of the labels
    $\hat{y}$ to the true labels $y$ through a cost function. See text
    for further explanation.}
    \label{training_fcnn}
    \begin{algorithmic}[1] % The number tells where the line numbering should start
      \Require choice of model hyperparameters
      \Require randomly initialized weights and biases
      \Procedure {Train}{$\bm{X}, \bm{Y}$} \Comment{input all training example features and labels}
      \Repeat %\Comment{optimize}
          \State $\mathcal{L} \gets 0$ \Comment{set cost to 0}
          \For {all $\bfx, y$ in $\bm{X}$, $\bm{Y}$} \Comment{consider each training example}
            \State $\hat{y} \gets \textrm{ForwardPass}(\bfx, \bfW, \bfb)$
              \Comment{predict with a forward pass through the network}
            \State $\mathcal{L} \gets \mathcal{L} + \textrm{Cost}(y, \hat{y})$
            \Comment{sum the cost of misfit to all data}
          \EndFor
          \State $\bfW, \bfb \gets \textrm{Update}(\textrm{Optimizer},\mathcal{L})$
            \Comment{update model based on total cost}
        \Until{model is optimized}
        \State \textbf{return} $\bfw, \bfb$ \Comment{return optimized weights and biases}
      \EndProcedure
    \end{algorithmic}
\end{algorithm}
\end{minipage}
\vspace{2\baselineskip}
}

\iffalse
\centerline{%

\begin{minipage}{0.9\textwidth}
\begin{algorithm}[H]
  \caption{Training a neural network. A set of example features
    $\bm{X}$ and corresponding labels $\bm{Y}$ are used to iteratively
    improve the model weights $\bfw$ and biases $\bfb$ from their
    initial random values by comparing model predictions of the labels
    $\bm{\hat{y}}$ to the true labels $\bm{y}$ through a cost
    function. See text for further explanation.}
    \label{training_fcnn}
    \begin{algorithmic}[1] % The number tells where the line numbering should start
      \Require choice of model hyperparameters
      \Require randomly initialized weights and biases
      \Procedure {Train}{$\bm{X}, \bm{Y}$} \Comment{input all training example features and labels}
        \Repeat %\Comment{optimize}
          \For {all $\bfx, \bm{y}$ in $\bm{X}$, $\bm{Y}$} \Comment{consider each training example}
            \State $\bm{\hat{y}} \gets \textrm{ForwardPass}(\bfx, \bfW, \bfb)$
              \Comment{predict with a forward pass through the network}
            \State $\mathcal{L} \gets \textrm{Cost}(\bm{y}, \bm{\hat{y}})$
              \Comment{cost of misfit to data}
          \EndFor
          \State $\bfW, \bfb \gets \textrm{Update}(\textrm{Optimizer},\mathcal{L})$
            \Comment{update model based on cost}
        \Until{model is optimized}
        \State \textbf{return} $\bfw, \bfb$ \Comment{return optimized weights and biases}
      \EndProcedure
    \end{algorithmic}
\end{algorithm}
\end{minipage}
\vspace{2\baselineskip}
}
\fi

Algorithms \ref{forward_pass} and \ref{training_fcnn} were coded in
the Python programming language and the neural networks were
implemented using the PyTorch imperative deep learning library
\citep[][]{pytorch:2019} on a Linux computer with an integral graphics
processing unit (GPU).

In the examples provided below, in lieu of the penalty approach for
regularization mentioned in Section \ref{sec:ann_regularization} ,
we regularized the learning problem with procedures provided with the
PyTorch software.  In particular, we invoked the built-in ``dropout''
or ``weight decay '' mechanisms \citep[e.g.][]{Goodfellow:2016,
  pytorch:2019}, applied within the gradient descent iteration, both
of which essentially detect and remove (set to zero) poorly
constrained components of $\bfgamma$.

To apply this model to the geothermal resource potential problem, the
10 feature maps from the Nevada Geothermal Play Fairway Project
(Figure \ref{fig_PFA_numerical_features}) were initially reformatted
into a flattened table where each column represents the feature
values, and these are organized row-wise keyed to the geographic
location on the map. The features were then prescaled so that they
each have a mean value of 0.0 and a standard deviation of 1.0.

The features of the benchmark sites (Figure \ref{fig_PFA_study_area}
inset) were extracted from this table for training and testing. There
are 62 positive sites and 84 negative sites in this database. The
training data set was balanced by randomly selecting 62 of the
negative sites from the original 84 examples. We found that with this
small number of training data the optimization was found to be erratic
in each training cycle and from one trial to the next. Therefore, the
data set was augmented for numerical stability by adding the 4 nearest
neighbors on the map to each true benchmark. These 620 examples in
total provided stability and repeatability to the training process.
The augmented benchmarks were randomly split into 67\% training
examples and 33\% testing examples. The models were then trained in
batches using back propagation and the Adam optimizer
\citep[e.g.][Chapter 8]{Goodfellow:2016}.

The learning curves, including the loss versus epoch and the training
and testing accuracy vs epoch, were used to gauge the process and used
to help choose the regularization hyperparameters to prevent
overfitting as seen by a large discrepancy between training and
testing accuracy. Both dropout and weight decay were explored as
regularizing methods.

Finally, once a model is successfully trained and tested and deemed
suitable, the probability of finding new positive sites in the study
area is predicted by feeding the features forward through the model at
each remaining geographic location (pixels). From these predictions a
geothermal resource map can be produced analogous to the original
Nevada Geothermal Play Fairway Project map \citep[Figure 1 of][]{Faulds:2018}.

\subsection{Results}

When we perform this training and prediction exercise and produce
typical outcomes for the resource maps, we notice two related issues
requiring further consideration.  First, there is the general issue of
confidence in the predictions driven by the need to choose additional
model heuristics related to regularization. The goal of this work is
to produce practical resource maps guiding geothermal exploration and
drilling. Confidence in the map is a key input to this decision making
process. A second and related issue is that even once all model
parameters are fixed, there is still a noticeable degree of variability
in the outcomes, especially in the details of the maps produced by
prediction from multiple optimizations of the model. This variability
must be understood and reconciled for the results to be most
useful. Below, we illustrate these issues with examples and then
propose a means to incorporate them into our understanding.

\subsubsection{Discussion of confidence in predictions}

A hidden aspect in all neural network models is the ability and need
to regularize the behavior in training in order to reduce over-fitting
to the training examples and ensure generalization to unseen
examples studied during the testing stage. In our network we have some
choices in this regard, first is to restrict any one weight or bias
from becoming too large and dominating the solution. This in PyTorch
is known as ``weight decay.'' Another method we can use, either
separately or combined with weight decay is to use ``dropout.'' This
works by randomly ignoring weights within the network with a set
probability so that the final solution is not too dependent on any
given link in the computation graph. Both of these methods impede the
ability of the network to overfit or indeed to ``memorize'' the
training examples.

\begin{figure}[p]
  \centering
  \includegraphics[width=0.8\textwidth]{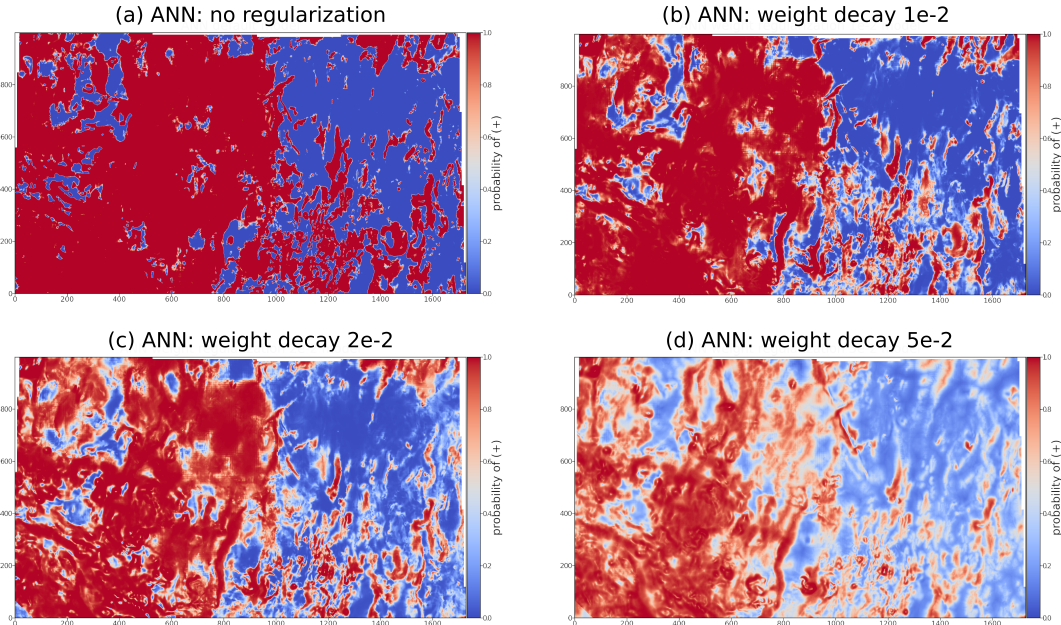}
  \caption{ANN model: illustration of reducing overfitting and
    overconfidence with regularization. Plotted are maps of the
    probability $p=P(\bfx,\bfgamma)$ predicted by the neural
    network. (a) Shows the optimized model without
    regularization. (b-c) show the effect of stronger regularization
    as controlled by increasing the weight decay parameter. For
    reference, these maps and those that follow overlay the gray area of
    Figure \ref{fig_PFA_study_area}, where north/south is aligned
    approximately up/down on the page and the axis units represent the
    indexes of the $250\textrm{m} \times 250\textrm{m}$ square
    pixels. See text for further discussion.
  } \label{fig_ANN_regularization}
\end{figure}

Finding optimal parameters or heuristics for regularization allows
training and testing accuracies to track each other and ensures the
model generalizes well. As an example, Figure
\ref{fig_ANN_regularization} illustrates 4 cases of different weight
decay settings and the large variability in the final geothermal
resource probability maps. If regularization is not strong (Figure
\ref{fig_ANN_regularization}a), the model gives the unsatisfactory
result that half of the study area has nearly a 100\% chance of being a viable
geothermal energy prospect. If more regularization is applied (for
example Figure \ref{fig_ANN_regularization}d) then much more nuance is
seen and the map becomes more believable and resembles the original
fairway predictions in Figure 1 of \citet[][]{Faulds:2018}. Clearly,
if we are uncertain as to how best to choose the regularization
parameters, then that translates to a concomitant uncertainty in the
result. We need means to quantify this effect.

\subsubsection{Discussion of variability in predictions}

Suppose that we have a way to choose the ``best'' regularization
method such that all model parameters are fixed. We find now that we have
a remaining variability in the predictions of the network when it is trained
multiple times. Training of artificial neural networks begins with a
starting set of weights and biases, usually set at random, and the
optimizer modifies these throughout the training stage toward a set of
optimum values.

\begin{figure}[p]
  \centering
  \includegraphics[width=0.8\textwidth]{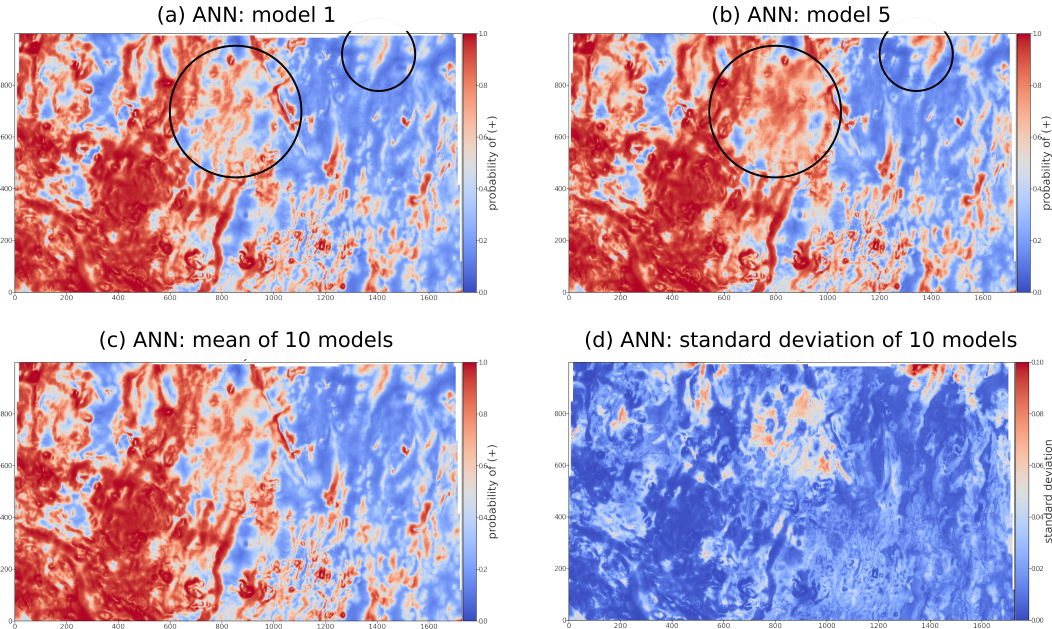}
  \caption{ANN model variability exposed by multiple runs, each having
    a different random number seed. Plotted are maps of the
    probability $p=P(\bfx,\bfgamma)$ predicted by the neural
    network. ANN models \#1 (a) and \#5 (b) represent the extreme
    variability seen among 10 training trials. A comparison of the two
    extremes to the mean (c) and the standard deviation (d) of all 10
    trials shows that there is significant variability due to random
    effects from one training trial to another. Note extreme examples
    within the circular regions.  } \label{fig_ANN_variability}
\end{figure}

For complex problems it is often the case that there is a highly
nonlinear relationship among the features and the outcome, the
features may not comprise a complete description of the physical
process, and the features may be further contaminated by measurement
noise. In this case there may not be an easily discoverable global
minimum in the cost function for the problem, but many local minima
may exist. Thus any training episode, given its random initial
configuration may not end up in the same place.  This is not easy to
see by simply watching the learning curves and by evaluating the
various performance metrics. It does become apparent for this problem
when predictions are made to produce geothermal resource maps in
unknown regions of the study area. Figure \ref{fig_ANN_variability}
illustrates this problem well.

In this example, the regularizing parameters of Figure
\ref{fig_ANN_regularization}d were set and 10 optimizations were
performed, each producing a resource map. Variability among trials
arises from random seeding of the weights and biases within the
network. Figure \ref{fig_ANN_variability}a and
\ref{fig_ANN_variability}b show two such maps with example regions
highlighted with circles. One can see noticeable differences in the
map details within these regions, yet it is not obvious everywhere
that they are different. Figure \ref{fig_ANN_variability}c shows the
average of 10 such maps, and Figure \ref{fig_ANN_variability}d shows
the standard deviation among these maps. It seems that some geographic
areas are simply more prone to variability than others, leading us to
the conclusion that the uncertainty varies significantly
spatially. This needs to be reconciled for confident application of
the results.

In what follows we describe a modification of the neural network model
that incorporates both of these forms of uncertainty and provides
useful quantitative measures of confidence and variability.

\section{Variational Inference with Bayesian Neural Networks}

\subsection{Paradigm}

As we have seen, the direct application of artificial neural networks
to the problem of geothermal resource assessment warrants further
analysis of model uncertainty. First, we wish to quantify the
variability in the state of the ``best'' model resulting from
seemingly identical successive training efforts. Second, once a model
is chosen we wish to define useful confidence bounds for predictions
arising during prediction. Together these issues expose aspects of
\textit{aleatoric} and/or \textit{epistemic} uncertainty.
\citet[][]{Hullermeier:2021} provide a lucid description of these
concepts in the context of machine learning models. We summarize some
of their key points here.

Machine learning involves the extraction of reasonable models from
data by example. The goal is to generalize from the cases presented to
the algorithm during training to new and yet unseen examples from
the data generating (physical) process. The resulting models can never
be proven to be correct, but remain hypothetical and uncertain.

Aleatoric, or statistical, uncertainty refers to the variability in
the outcome due to random effects. In this application the following
sources of randomness are related to feature input data: inherent
randomness in geophysical and geoglogic features, measurement error,
and interpolation or smoothing between geographic measurement
points. There are also model related sources of variability,
including: random initialization of weights, dropout regularization,
incomplete convergence due to early termination of training, and
convergence to a one of perhaps many local minima in a complex cost
function. These sources of uncertainty are considered ``irreducible'' in
that they cannot in practice be reduced to an arbitrary desired level.

Epistemic uncertainty represents the lack of knowledge about the best
model to use by the decision maker and not to
any underlying random process. This type of uncertainty can be reduced
in principle through incorporating additional information. Such
information could include the addition of further training examples,
augmentation of the feature set through engineering the features
themselves or adding new data types, or adding to or modifying the set
of model hypotheses considered in the search for the best model (such
as modifying the neural network architecture). These sources of
uncertainty are under our control and so epistemic uncertainty refers
to the ``reducible'' part of the total.

Since we use artificial neural networks, epistemic uncertainty is
assumed to be uncertainty about the model parameters $\bfgamma$, that
is, the values of both the weights and biases
$\bfgamma = (\bfW, \bfb)$. Bayesian neural networks
\citep[][]{Denker:1991} represent a paradigm that captures this type
of uncertainty. We note, however, that the Bayseian neural network
approach does not explicitly recognize the dichotomy among nor attempt
to isolate different sources of uncertainty.

Using the Bayesian approach, instead of having a fixed set of
single-valued model parameters to be optimized, each weight and bias
in the network is represented by its own probability
distribution. Learning the parameters of these distributions comes
about in essence by the application of Bayes's rule
\citep[e.g.][]{Bishop:2006}. Thus the posterior of the weight
distributions is computed as an update from our prior assumptions or
knowledge by considering the evidence from the training data.

The motivation for the use of Bayesian neural networks is that they
allow us to both naturally incorporate model variability and to
provide confidence bounds for our predictions. This approach includes
the epistemic uncertainty described above and also incorporates the
model related sources of variability listed in the aleatoric
category. We leave analysis of the data-related uncertainty in the
machine learning models for another time.

In the following sections, we develop this concept from theory to
practice. Beginning with a discussion of the classification problem, we
summarize the formalism for the use of Bayesian neural networks (BNN) and in
so doing we underscore the meaning of the outputs of these
models. Finally, we provide notes on their implementation and
optimization and finish with the application to geothermal resource
assessment.

%%%%%%%%%%%%%%%%%%%%%%%%%%%%%%%%%%%%%%%%%%%%%%%%%%%%%%%%%%
% BEGIN Bill's 2nd Section
%%%%%%%%%%%%%%%%%%%%%%%%%%%%%%%%%%%%%%%%%%%%%%%%%%%%%%%%%%

\subsection{Mathematical Formulation for BNN}

\subsubsection{Bayesian \NN}
\label{sec:bnn}

Reiterating from Section \ref{sec:mf_ann}, the function $P$ in
\eqeqp{Pdef} represents a deterministic \nn\ that predicts
the proportion of positive sites among a population of sites
characterized by a feature vector $\bfx$.  The second
argument of $P$, the vector $\bfgamma$, represents the
network parameters (weights and biases).  A Bayesian \nn, in
contrast, predicts a \pdf\ (PDF) of $p$ based on a specified
PDF of $\bfgamma$, again for given $\bfx$.  Denoting the PDF
of $\bfgamma$ as $f(\bfgamma)$, a Bayesian network evaluates
\begin{align}
f(p\cnd\bfx) &= \int \abd\bfgamma\ f(p\cnd\bfx,\bfgamma)\,f(\bfgamma)
  \labeq{pdfpbarx}
\end{align}
where $\abd\bfgamma=\dee\gamma_1\, \dee\gamma_2\, \cdots$
denotes the differential volume in $\bfgamma$-space and
where the integration is over the entire space.  This
equation follows from the definitions of marginal and
conditional probability distributions and the requirement
that the PDF of $\bfgamma$ not depend on $\bfx$,
i.e., $f(\bfgamma\cnd\bfx)=f(\bfgamma)$.  The deterministic
function $P$ underlies this calculation in that \eqp{Pdef}
implies
\begin{align}
f(p\cnd\bfx,\bfgamma) &= \delta(p-P(\bfx,\bfgamma))
  \labeq{pdfpbarxgam}
\end{align}
$\delta$ being the Dirac delta function.

We note that, even for simple choices of $f(\bfgamma)$,
\eqp{pdfpbarx} cannot in general be reduced to an analytic
expression given that $P$ is a nonlinear function of
$\bfgamma$.  Therefore, our implementation of Bayesian
networks represents $f(p\cnd\bfx)$ indirectly as a set of
random samples drawn from this PDF: $p_k$, $k=1$, 2, \dots.
These are generated as $p_k=P(\bfx,\bfgamma_k)$ where each
$\bfgamma_k$ is a random samples from $f(\bfgamma)$.  The
samples $p_k$ can be used to construct a histogram that
serves as an approximation to $f(p\cnd\bfx)$, and to
estimate moments or quantiles of $f(p\cnd\bfx)$.

Given $f(p\cnd\bfx)$, one can derive the PMF of the discrete
random variable $y$ as follows.  First we express the PMF as
(similar to \eqp{pdfpbarx})
\begin{align}
f(y\cnd\bfx) &= \int_0^1\dee p\ f(y\cnd p,\bfx) \, f(p\cnd\bfx).
  \labeq{pmfybarx_bnn}
\end{align}
Substituting the Bernoulli model of \eqseqp{bern} and
\eqp{phidesx} we have
\begin{align}
f(y\cnd\bfx) &= \int_0^1\dee p\ p^y \, (1-p)^{1-y} \, f(p\cnd\bfx).
  \labeq{pmfybarx_bnn2}
\end{align}
Evaluating this integral separately for $y=0$ and 1, then
recombining, we obtain
\begin{align}
f(y\cnd\bfx) &= \Pbar(\bfx)^y \, \big(1-\Pbar(\bfx)\big)^{1-y}
  \labeq{pmfybarx_bnn3}
\end{align}
where
\begin{align}
\Pbar(\bfx) &= \int_0^1 \dee p\ p \, f(p\cnd\bfx).
  \labeq{Pbar}
\end{align}
We see that $y$ still follows a Bernoulli distribution, with
the probability of success given by the mean of
$f(p\cnd\bfx)$.

\subsubsection{Training the Bayesian Network}

With a Bayesian \nn\ the training data are used to learn the
PDF $f(\bfgamma)$, as opposed to learning a single,
best-fitting value of $\bfgamma$.  True to its name, a
Bayesian network enlists Bayesian inference for this task.
That is, $f(\bfgamma)$ is taken to be the posterior PDF of
$\bfgamma$, which we denote $f_\pos(\bfgamma)$, as informed
by the training data.  Applying Bayes's rule we have
\begin{align}
f_\pos(\bfgamma)
&\equiv f(\bfgamma\cnd X,Y)
= \frac
  { f(Y\cnd X,\bfgamma) \, f_\pri(\bfgamma) }
   { f(Y\cnd X) }
%   { \int \abd\bfgamma\
%      f(Y\cnd X,\bfgamma) \, f_\pri(\bfgamma) }
  \labeq{bayess_rule}
\end{align}
where $f_\pri(\bfgamma)$ is an assigned prior PDF of
$\bfgamma$, $f(Y\cnd X,\bfgamma)$ is the likelihood function
given by \eqeqp{like}, and
\begin{align}
f(Y\cnd X) &= \int \abd{\bfgamma}\ f(Y\cnd X,\bfgamma) \, f_\pri(\bfgamma)
\end{align}
which is the PDF of $Y$ marginalized \wrt\ $\bfgamma$.

As we noted in Section \ref{sec:bnn}, the direct computation
of $f(p\cnd\bfx)$ is in general not feasible and certainly
not when the PDF of $\bfgamma$ is taken to be the Bayesian
posterior, $f_\pos$ in \eqp{bayess_rule}.  The alternative
of generating random samples from $f(p\cnd\bfx)$ also faces
computational challenges in requiring the sampling of
$f_\pos(\bfgamma)$.  To accommodate this requirement, we
adopted the variational Bayes method to obtain an
approximation to $f_\pos$ amenable to efficient sampling.

\subsubsection{Variational Bayes}

In the variational Bayes method $f_\pos(\bfgamma)$ is
approximated with a convenient representation involving a
vector of parameters, $\bftheta$: $f_\pos(\bfgamma) \approx
g(\bfgamma;\bftheta)$.  In this study we took $g$ to be a
multivariate Gaussian p.d.f.\ with the components of
$\bftheta$ comprising the means and variances of the
components of $\bfgamma$ (the \nn\ biases and weights).  The
covariances between network parameters were assumed to be
zero, reducing the sampling of $g(\bfgamma;\bftheta)$ to the
easy task of sampling univariate Gaussian distributions.

Variational Bayes seeks the value of $\bftheta$ that minimizes the
discrepancy between $g$ and $f_\pos$, as quantified by the
Kullback-Leibler (KL) divergence of $g$ from
$f_\pos$. \citet[][Chapter 3]{Goodfellow:2016} provide further context
for use of the KL divergence in terms of Information Theory. In our
notation it is defined as:
\begin{align}
\DKL[g || f_\pos] &=
  \int \abd\bfgamma\ g(\bfgamma;\bftheta) \,
    \log \left( \frac {g(\bfgamma;\bftheta)} {f_\pos(\bfgamma)} \right)
  \labeq{KL}
\end{align}
where the integration is performed over the support of $g$,
i.e.\ the regions of $\bfgamma$-space in which $g>0$.  We
recognize this as an expected value \wrt\ $\bfgamma$,
using $g$ as its PDF.  For any function of $\bfgamma$,
$\phi(\bfgamma)$, let us define
\begin{align}
\E_g\big[\phi(\bfgamma)\big] &=
  \int \abd\bfgamma\ g(\bfgamma;\bftheta) \, \phi(\bfgamma).
\end{align}
We can then write \eqp{KL} as
\begin{align}
\DKL[g || f_\pos] &=
\E_g\left[  \log \left(
  \frac {g(\bfgamma;\bftheta)} {f_\pos(\bfgamma)} \right) \right].
  \labeq{KL2}
\end{align}
Plugging in \eqeqp{bayess_rule} (Bayes's rule) and expanding
the logarithm, we get
\begin{align}
\DKL[g||f_\pos] &=
\E_g \left[
    \log \left( \frac {g(\bfgamma;\bftheta)\,f(Y\cnd X)}
    {f(Y\cnd X,\bfgamma)\,f_\pri(\bfgamma)} \right) \right]
\nonumber \\
&=
- \E_g \Big[ \log f(Y\cnd X,\bfgamma) \Big]
+ \E_g \left[
    \log \left(
      \frac {g(\bfgamma;\bftheta)}
            {f_\pri(\bfgamma)} \right) \right]
+ \log f(Y\cnd X).  \labeq{KL3}
\end{align}

We recognize the first term on the \rhs\ of \eqp{KL3} as the
expected value of the negative log-likelihood function,
$\calL$ in \eqeqp{calL}, and the second term as the
KL divergence of $g$ from the {\em prior} distribution
$f_\pri$.  Since the third term does not depend on
$\bftheta$, we can thus express the variational Bayes
solution for $\bftheta$ as
\begin{align}
\bftheta_\VB &= \arg\min_\bftheta \calK(\bftheta;X,Y)
  \labeq{thetaVB}
\end{align}
where the objective, or cost, function $\calK$ is given by
\begin{align}
\calK(\bftheta;X,Y) &= 
  \E_g \calL(\bfgamma;X,Y) + \DKL[g||f_\pri].  \labeq{calK}
\end{align}
We observe that the first term of $\calK$ is the expected
value of the cost function minimized in deterministic
network training (to obtain $\bfgamma_\ML$), while the
second term serves as a penalty function that regularizes
$\bftheta_\VB$.

We use a gradient descent method to minimize the cost
function $\calK$.  The algorithm is similar to that used to
calculate $\bfgamma_\ML$ for a deterministic \nn\ but with
some notable differences.  In particular, since $\calK$
includes a penalty term, we do not invoke regularization
procedures (like dropout) within the gradient descent loop.
Further, evaluation of the cost function and its gradient
now entails finding an expected value \wrt\ $\bfgamma$, by
virtue of the first term of \eqp{calK}.  This is
accomplished by averaging $\calL$ and its gradient over
values of $\bfgamma$ randomly sampled from
$g(\bfgamma;\bftheta)$, with $\bftheta$ fixed to its current
value at each step of the descent. Further details of our
computational approach are given below in Section
\ref{sec:implement_section}.

\subsubsection{Regularization Parameter}

Within the Bayesian inference paradigm, the cost function
$\calK$ in \eqp{calK} optimally pools the information
provided by the training data and by prior knowledge.
However, this optimality assumes that the statistical models
for the two sources of information are equally credible and
well calibrated, which is not the case in the application
presented here.  In particular, the prior distribution
assigned to \nn\ parameters, $f_\pri(\bfgamma)$, is at best
conjectural.  The formal remedy for such deficiencies is to
introduce unknown hyperparameters into the statistical
models for the training data and prior information.  The
selection of these parameters is addressed within the
Bayesian inference paradigm with the goal of balancing the
two sources of information to avoid overfitting or
underfitting the training data.

In this project we applied a simple version of this remedy
by introducing a single hyperparameter, $\alpha>0$, and
replacing the cost function in \eqeqp{calK} with
\begin{align}
\calK(\bftheta;X,Y) &= 
  \E_g \calL(\bfgamma;X,Y) + \alpha \, \DKL[g||f_\pri].  \labeq{calKalf}
\end{align}
By scaling the penalty term, $\alpha$ plays the role of a
regularization parameter as used in inversion methods like
Tikhonov regularization.  We note that in some problems a
regularization parameter can be identified with a parameter
of the underlying statistical model, such as the variance of
a measurement error or a prior variance of the unknown
parameters.  In our variational Bayes application, with the
Bernoulli model assumed for training data, $\alpha$ does not
have such a simple interpretation.
In Section \ref{sec:modeling} we suggest that
$\alpha$ controls the model complexity and following this notion we
propose a practical numerical procedure for selecting it.

\subsubsection{Interpretation of $f(p\cnd\bfx)$}

Our classification framework bases the assessment of a
prospective geothermal site with feature vector $\bfx$ on
the PDF $f(p\cnd\bfx)$, as calculated by the Bayesian
\nn\ trained with data from known positive and negative
sites.  The parameter $p$ is the proportion of positive
sites among all sites sharing feature vector $\bfx$.  Its
PDF quantifies our uncertainty about its true value and can
be used, for example, to construct confidence intervals (or
credibility intervals) on $p$, or to test hypotheses about
$p$.

The goal of the classification problem, however, is to predict
the class of a site, $y$, based on its PMF, $f(y\cnd\bfx)$.
Owing to the simplicity of the Bernoulli model, our analysis
concluded that $f(y\cnd\bfx)$ depended only on the mean of
$f(p\cnd\bfx)$ and not its shape.  Effectively, uncertainty
in the value of $p$ does not affect uncertainty in $y$.
Yet, intuitively, the shape of $f(p\cnd\bfx)$ should play a
role in deciding how to classify a site.  In particular, it
would be reasonable to consider the dispersion of
$f(p\cnd\bfx)$ about the mean --- the width of the
distribution --- in deciding whether or not a site is
positive.  For a given mean, a smaller dispersion should
make one feel more confident in the class assigned to a
site.

While this seeming deficiency in our analysis might be
remedied with a more sophisticated statistical model, it
partly stems from the fundamental meaning of probability.  A
probability predicts the average outcome of a repeated
random experiment over the long run.  In our problem, an
experiment can be viewed as a two-step process: first select
$p$ at random from $f(p\cnd\bfx)$, and then draw a site at
random from the population of sites
that
have a proportion $p$ of
positive sites.  Our analysis showed that, over the long
run, the proportion of outcomes with $y=1$ (positive sites)
will be $\pbar$, the mean of $f(p\cnd\bfx)$.  Outcomes where
the first step yields $p<\pbar$ will negate those yielding
$p>\pbar$, regardless of the shape of $f(p\cnd\bfx)$.

However, a decision maker concerned about false positives
(classifying negative sites as positive) might want to
perform a ``worst case,'' rather than average case, analysis
and consider the following two-step experiment: randomly
select $K$ (say, ten) values of $p$ from $f(p\cnd\bfx)$ and
then draw a site at random from the population having the
{\em smallest} $p$ among the $K$ samples.  Pursuing this
idea, let $F(p\cnd\bfx)$ be the \cdf\ (CDF) associated with
$f(p\cnd\bfx)$:
\begin{align}
F(p\cnd\bfx) &= \int_0^p \dee p'\ f(p'\cnd\bfx).
\end{align}
Using integration by parts, one can verify that
\begin{align}
\pbar &\equiv
  \int_0^1 \dee p\ p\,f(p\cnd\bfx)  
  = \int_0^1 \dee p\ \big(1- F(p\cnd\bfx)\big).
  \labeq{pbarvsF}
\end{align}
Now denote the PDF and CDF of the worst-of-$K$ version of
$p$ as $f_\worst$ and $F_\worst$, respectively.  A basic
result of order statistics \citep[e.g.][]{David:1981} is
\begin{align}
1 - F_\worst(p\cnd\bfx) &= \big(1 - F(p\cnd\bfx)\big)^K.
\end{align}
It follows that the mean of $f_\worst(p\cnd\bfx)$, which
becomes the inferred probability of a positive site in this
worst-of-$K$ analysis, is
\begin{align}
\pbar_\worst &=
  \int_0^1 \dee p\ \big(1- F(p\cnd\bfx)\big)^K.
  \labeq{pworstbarvsF}
\end{align}
%Readers can convince themselves
It happens that $\pbar_\worst$ is less
than $\pbar$ and does depend on the shape of $f(p\cnd\bfx)$,
with $\pbar_\worst$ deceasing as the dispersion of
$f(p\cnd\bfx)$ increases.

To acknowledge the role uncertainty in $p$ plays in site
classification, without committing to a particular decision
strategy, the results of the next section display a quantile
of $f(p\cnd\bfx)$ rather then its mean.  Namely, we show the
value of $p$ satisfying
\begin{align}
F(p\cnd\bfx) &= \beta
\end{align}
for some choice of $\beta$.  A neutral choice would be
$\beta=0.5$ but we have chosen $\beta=.05$ to reflect a
greater aversion to false positive classifications than
false negative classifications.

%%%%%%%%%%%%%%%%%%%%%%%%%%%%%%%%%%%%%%%%%%%%%%%%%%%%%%%%%%
% END Bill's 2nd Section
%%%%%%%%%%%%%%%%%%%%%%%%%%%%%%%%%%%%%%%%%%%%%%%%%%%%%%%%%%

\subsection{Implementation}
\label{sec:implement_section}

The implementation of a Bayesian neural network involves several key
changes to that of the deterministic fully connected network described
earlier in Algorithms \ref{forward_pass} and
\ref{training_fcnn}. Before, the network model was determined by a
single set of deterministic parameters (weights and biases).  In the
Bayesian case the network is determined by distributions of weights
and biases, in effect representing an infinite family of models. For
practical purposes each parameter distribution is assumed to be
Gaussian, parameterized by its mean and standard deviation. Thus, the
number of parameters defining a Bayesian neural network is twice that
of a deterministic neural network with the same architecture.

When the model is used for prediction, a specific set of weights and
biases must first be sampled from the distributions for each node in
the network. The key elements of sampling from the posterior
distributions of the weights and biases are described in Algorithm
\ref{sample_posteriors}. Then, once this posterior sampling is
performed, a fixed set of weights and biases is in hand and can then
be used to generate a prediction using the same algorithm as in the
standard network (Algorithm \ref{forward_pass}).

\centerline{%
\begin{minipage}{0.9\textwidth}
\begin{algorithm}[H]
  \caption{Sampling from the posterior distributions of weights and
    biases for a Bayesian neural network. The network is defined
    probabilistically with normal distributions of weights and
    biases. Each of these distributions is parameterized by its mean
    $\mu$ and standard deviation $\sigma$. To facilitate training, the
    standard deviation is forced to be positive by using an
    intermediate real-valued parameter $\rho$ and then applying the
    PyTorch \textit{softplus} function
    \citep[e.g.][]{pytorch:2019}. Forward passes through the network
    first require sampling from these posterior distributions.}
    \label{sample_posteriors}
    \begin{algorithmic}[1] % The number tells where the line numbering should start
      \Procedure {SamplePosteriors}{$\bm{\rho}, \bm{\mu}$}
        \For {all weight elements i} \Comment{weight matrices}
          \State $\sigma^{[w]}_{i} \gets \log ( 1 + \exp(\rho^{[w]}_{i}) )$ 
          \Comment{create a positive standard deviation with \textit{softplus}}
          \State $w_{i} \gets \mathcal{N} (0,1) \cdot \sigma^{[w]}_{i} + \mu^{[w]}_{i}$
          \Comment{sample from a standard normal distribution and scale}
        \EndFor
        \For {all bias elements j} \Comment{bias vectors}
          \State $\sigma^{[b]}_{j} \gets \log ( 1 + \exp(\rho^{[b]}_{j}) )$
          \Comment{create a positive standard deviation with \textit{softplus}}
          \State $b_{j} \gets \mathcal{N} (0,1) \cdot \sigma^{[b]}_{j} + \mu^{[b]}_{j}$
          \Comment{sample from a standard normal distribution and scale}
        \EndFor
        \State \textbf{return} $\bfW$, $\bfb$ \Comment{return weights and biases}
      \EndProcedure
    \end{algorithmic}
\end{algorithm}
\end{minipage}
\vspace{2\baselineskip}
}

The training of a Bayesian neural network contains some further
modifications from the deterministic case as described in Algorithm
\ref{training_bnn}. First, the cost function being optimized
(minimized) is now $\calK$ in Equation \eqp{calKalf}.  For each training
iteration the algorithm employs a sampling of the posteriors before
predictions are made. This is an important nuance, since the
likelihood term is now an expected value (i.e. $\E_g$) with respect to
the posterior of $\bfgamma$. Since each posterior sampling and forward
pass is merely a sample from the model set then we must gather some
statistics of the properties of the model in order to meaningfully update
the mean and standard deviation of each weight and bias
distribution. Therefore, several realizations of the model are probed
and accumulated -- first using posterior sampling, then a forward pass, and then
computation of the cost -- before updating the statistics (means and
standard deviations) of the weight and bias distributions.  Finally,
we note that in training a Bayesian network the cost function is now
composed of two terms, one a penalty for model complexity and the other a
penalty for misfit to the training data (Equation
\eqp{calKalf}). The balance between these two terms is determined by
an important heuristic parameter, $\alpha$. A suitable choice of this
heuristic is described in the next section.

\centerline{%
\begin{minipage}{0.9\textwidth}
\begin{algorithm}[H]
    \caption{Training a Bayesian neural network (see text for explanation).}
    \label{training_bnn}
    \begin{algorithmic}[1] % The number tells where the line numbering should start
      \Require choice of prior distributions for weights and biases
      \Require choice of cost function heuristic parameter, $\alpha$
      \Require choice of model hyperparameters
      \Require randomly initialized posterior parameters for weights and biases, $\rho$ and $\mu$ 
      \Require choice of number of samples for gathering posterior statistics, $N_{samples}$
      \Procedure {Train}{$\bm{X}, \bm{Y}$} \Comment{input all training example features and labels}
      \Repeat %\Comment{optimize}
          \State $\mathcal{L} \gets 0$ \Comment{set cost to 0}
          \For {all $\bfx, y$ in $\bm{X}$, $\bm{Y}$} \Comment{consider each training example}
            \For {$j$ in $N_{samples}$} \Comment{gather statistics of posteriors}
              \State $\bfW_j, \bfb_j \gets \textrm{SamplePosteriors}(\bm{\rho}, \bm{\mu})$
                \Comment{sample the posterior distributions}
              \State $\hat{y}_j \gets \textrm{ForwardPass}(\bfx, \bfW_j, \bfb_j)$
                \Comment{predict with a forward pass through the network}
              \State $\mathcal{L}_j \gets \textrm{Cost}(y, \hat{y}_j,
                      \textrm{priors}, \bm{\rho}, \bm{\mu}, \alpha)$
                \Comment{cost balancing data misfit and model complexity}
            \EndFor
            \State $\mathcal{L}_{N} \gets \sum\mathcal{L}_j/N_{samples}$
              \Comment{Average cost for $N_{samples}$}
          \State $\mathcal{L} \gets \mathcal{L} + \mathcal{L}_{N}$ \Comment{sum the cost of misfit to all data}
          \EndFor
          \State $\bm{\rho}, \bm{\mu} \gets \textrm{Update}(\textrm{Optimizer},\mathcal{L})$
            \Comment{update model based on total cost}
        \Until{model is optimized}
        \State \textbf{return} $\bm{\rho}, \bm{\mu}$ \Comment{return optimized model parameters}
      \EndProcedure
    \end{algorithmic}
\end{algorithm}
\end{minipage}
\vspace{2\baselineskip}
}

\subsection{Modeling}
\label{sec:modeling}

We now have in hand a formulation of a Bayesian neural network
clarifying the interpretation of the outcomes and an algorithm for its
implementation on a computer. We next explore the properties of this
approach and compare them to the deterministic neural network
discussed earlier.

\subsubsection{Controlling the Model Complexity}

\begin{figure}[p]
  \centering
  \includegraphics[width=0.8\textwidth]{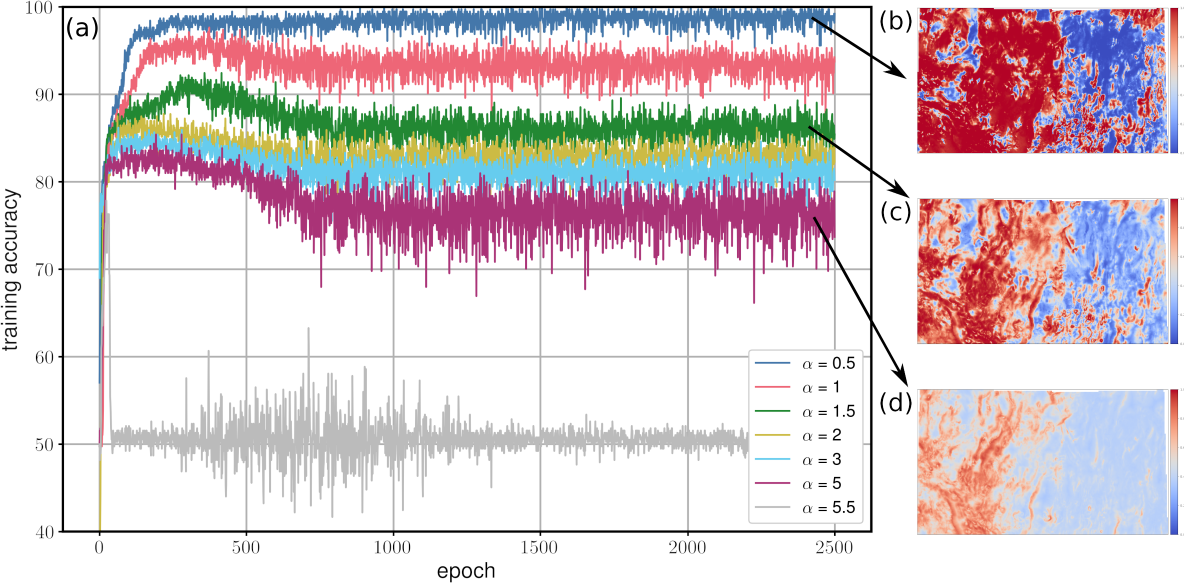}
  \caption{Illustration of the control of model complexity by varying
    the regularization parameter $\alpha$. (a) shows a sequence of
    learning curves tracking the evolution of the training accuracy
    with training epoch. As $\alpha$ increases the final training
    accuracy at epoch 2500 decreases systematically and moves from
    overfitting (where accuracy approaches 100\% at $\alpha = 0.5$) to
    not being able to classify any longer (where accuracy is about
    50\% at $\alpha =5.5$). The effect of varying $\alpha$ in this
    range is shown in (b-d) where more reasonable maps result at
    intermediate values of regularization.
  } \label{fig_BNN_alpha_training}
\end{figure}

In this discussion we refer to the cost function of Equation
(\refeq{calKalf}) which controls the balance between model complexity
and data fit by varying a regularization parameter $\alpha$. We
illustrate the effect of $\alpha$ by example in Figure
\ref{fig_BNN_alpha_training}.

The same feature data set and geothermal site labels used in the
previous examples were used to optimize a set of Bayesian models by
varying the regularization parameter $\alpha$. To illustrate this
process, we monitor the training accuracy learning curves along with
examples of the resulting predicted geothermal potential maps (see in
Figure \ref{fig_BNN_alpha_training}). Focusing on the learning curves,
we see that for small values of $\alpha$, where the data fit term in
the loss dominates, the model attains high accuracy. This is also true
for testing accuracy, but this is not shown here. As the
regularization parameter is increased, the training accuracy goes down
until at some point the algorithm fails to classify the examples and the
training accuracy abruptly drops to about 50\%.

We also show the maps of the study area produced by several examples
of the trained models. For the small $\alpha$ and high training
accuracy cases the geothermal resource maps are effectively binary in
character ... the model largely predicts either a positive geothermal
resource or a negative one with little in between. There is no nuance
in the predictions and an unrealistically large proportion of the study
area are predicted to be positive sites. These models are therefore not
useful.

When $\alpha$ increases, however, the maps have nuance in the
predictions, there is a much lower overall areal proportion of high
resource potential, and the maps more closely resemble the results
from the original PFA study. This all serves to illustrate the
importance of properly choosing the regularizing parameter in order to
obtain a useful and defensible result.

\subsubsection{An Optimal Degree of Regularization}

We conclude that the key to finding an optimal degree of regularization
is to consider the information content a particular model is capable
of holding. After a brief review of some literature on this subject we
will pose a first order and workable choice of the regularization
parameter based on this concept.

\paragraph{Background.}

We have seen the need to regularize the solution by varying the
parameter $\alpha$ in Equation (\refeq{calKalf}) that balances the
fit to the data and the complexity of the model. It is tempting to
treat this regularization process, not as a simple heuristic choice,
but instead tie it to a quantifiable measure defining a best solution.

Some inspiration can be derived from the literature describing
inversion via the maximum entropy principle. The works by Gull,
Skilling, and others \citep{Gull:1978, Gull:1989, Skilling:1989,
  Skilling:1990, Skilling:1991} provide good descriptions of this
concept and its application. In these papers a solution is sought that
contains no more structure than the experimental data ``allow.'' For
example, in an image processing problem, \citet{Gull:1989} considers a
heuristic similar to ours and identifies it with the model complexity
and further links it to the (hopefully) countable number of ``good'' or
accurate singular vectors contained in the data.

The concept of degrees of freedom is discussed lucidly by
\citet{Walker:1940}, where it is described as the number of free
parameters available to adjust a geometric figure in an N-dimensional
space. Of significance to our problem is the consideration of a
hyperplane, defined as any equation of the first degree connecting the
N variables. For example, a line in two-dimensional space is a
hyperplane, as is a plane in three-dimensional space, etc. A
hyperplane has N-1 degrees of freedom of movement on it and, in its
simplest form, N parameters to define it. The simplest decision
boundary one could seek in an N-dimensional classification problem
would be a hyperplane.

\citet{AbuMostafa:2012} summarize the definition and significance of
the Vapnik–Chervonenkis (or VC) dimension, $d_{VC}$, as measure of
model complexity.  This gives both an upper bound to the
``out-of-sample'' error and the ``effective'' number of degrees of
freedom of the model under consideration. This concept is said to
provide a means for choosing the best regularizing parameter, and using
the VC dimension in this way has the same goal as cross-validation,
the process of choosing a regularization parameter by ensuring optimal
generalization by using a hold-out data set
\citep[e.g.][]{Goodfellow:2016}. This concept holds for any model, be
it a neural network, support vector machine \citep[e.g.][Chapter
5]{Goodfellow:2016}, or other. An interesting end member case is that
of the perceptron, where the VC dimension is equal to one more than
the number of features, i.e. $d_{VC} = N+1$, a value close to the
number of parameters needed to specify a hyperplane.

Yet other attempts to balance the data fit and model complexity make
use of the Akaike Information Criterion (AIC) and the Bayesian
Information Criterion (BIC) and their variants
\citep[e.g.][]{AIC_BIC:2014}.  Such concepts of defining model
complexity and information content just mentioned and more have been
discussed widely in the literature in the search for the perfect
heuristic for choice of regularization. In consideration of artificial
neural networks some notable further examples are espoused by
\citet{Hinton:1993}, \citet{Fisher:2003}, \citet{Kline:2010}, \citet{
  Gao:2016}, and \citet{Maddox:2020}.

In principle, then, we are presented with the problem of finding a
model that is as simple as possible, that fits our training data, and
generalizes well to test data during cross validation and thereby
honoring the information available in the data set in the presence of
noise, measurement error, etc. Ideally, we also have some useful
metrics to aid us in this search. In practical problems the key
question is whether or not we can measure or estimate all of the
quantities required by the theories.

Guided by these concepts, and in light of our unaccounted-for possible
measurement errors and noise, in the following few sections we
describe our approach to find a conservative degree of regularization.

\paragraph{Counting Degrees of Freedom.}
An artificial neural network contains a fixed number of free
parameters (weights and biases) controlling the nonlinear combinations
of input features. Among the large number of parameters a model
contains, once the model is trained only a relatively small number
contribute significantly to the final predictions. One can imagine
determining the number of dominant model parameters by counting those
larger than some threshold. This technique has been used frequently to
``prune'' a network to make it more compact in terms of memory use by
setting weights smaller than a threshold to zero and perhaps
reconfiguring the architecture \citep[e.g.][]{Han:2015}. Choosing such
a threshold could be considered somewhat arbitrary.

For our purposes using Bayesian neural networks, the problem is more
straight forward. In the Bayesian case, the weights and biases are not
fixed values but are instead drawn form probability
distributions. These distributions are approximated by Gaussians and
are each parameterized by their means and standard deviations. The
useful realization is that if a particular distribution is too broad
and has a mean too close to zero, then random samples from the
distribution could take on both positive and negative values. If this
happens too frequently during prediction, then that particular weight
cannot possibly contribute effectively to the outcome as sometimes it
would add a contribution and another time it would subtract. We
therefore propose the following method for counting the effective
degrees of freedom of a Bayesian neural network.

We begin with a trained model, which has been optimized with a
particular choice of the regularization parameter $\alpha$. For this
trained model we inspect all of the weight distributions and sort them
by the absolute value of their ratios of the mean to the standard
deviation, $|\mu/\sigma|$. We count a weight as useful and
contributing to the number of degrees of freedom if
$|\mu/\sigma| \geq 1$. On the other hand if $|\mu/\sigma| < 1$ the
weight is assumed to not be effective in the computations since too
many random samples drawn from it are of alternating sign. In
general. such noncontributing weights will also necessarily have a
small value upon sampling, else they would simply add noise to the
predictions.

%%%%%%%%%%%%%%%% 
%\fi

We now show how this criterion can be applied to find a conservative
value of the regularization parameter.

\paragraph{Search for an Optimal Regularizer.}
\label{sec:optimal}

\begin{figure}[p]
  \centering
  \includegraphics[width=0.8\textwidth]{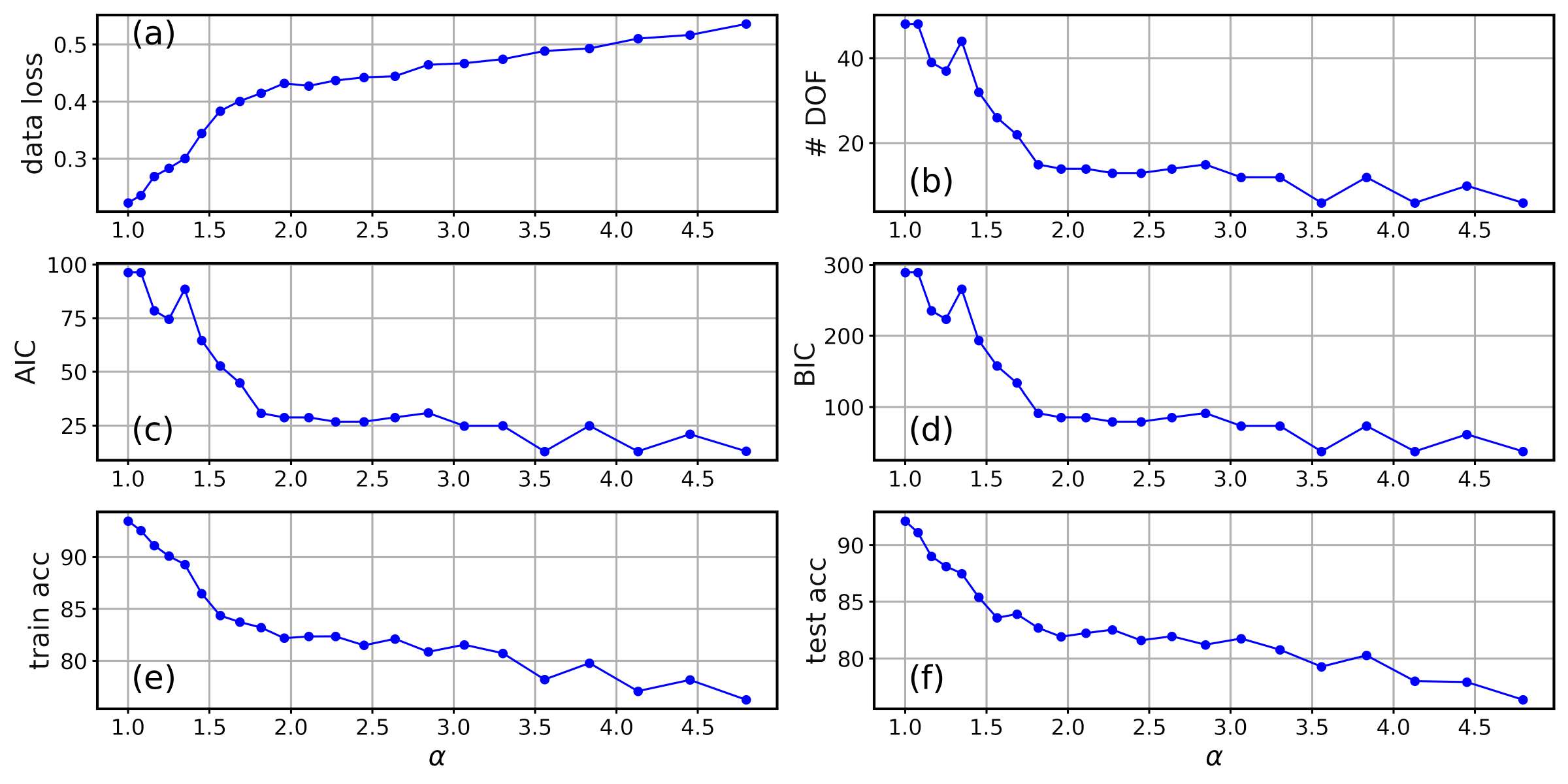}
  \caption{ Metrics tracking effect of varying the regularization
    parameter $\alpha$. We show (a) the data loss, (b) our estimate of
    the number of degrees of freedom (DOF), (c) the Akaike Information
    Criterion (AIC), (d) the Bayes Information Criterion (BIC), (e) the training
    accuracy, and (f) the testing accuracy.
  } \label{fig_BNN_alpha_loop}
\end{figure}

\begin{figure}[p]
  \centering
  \includegraphics[width=0.8\textwidth]{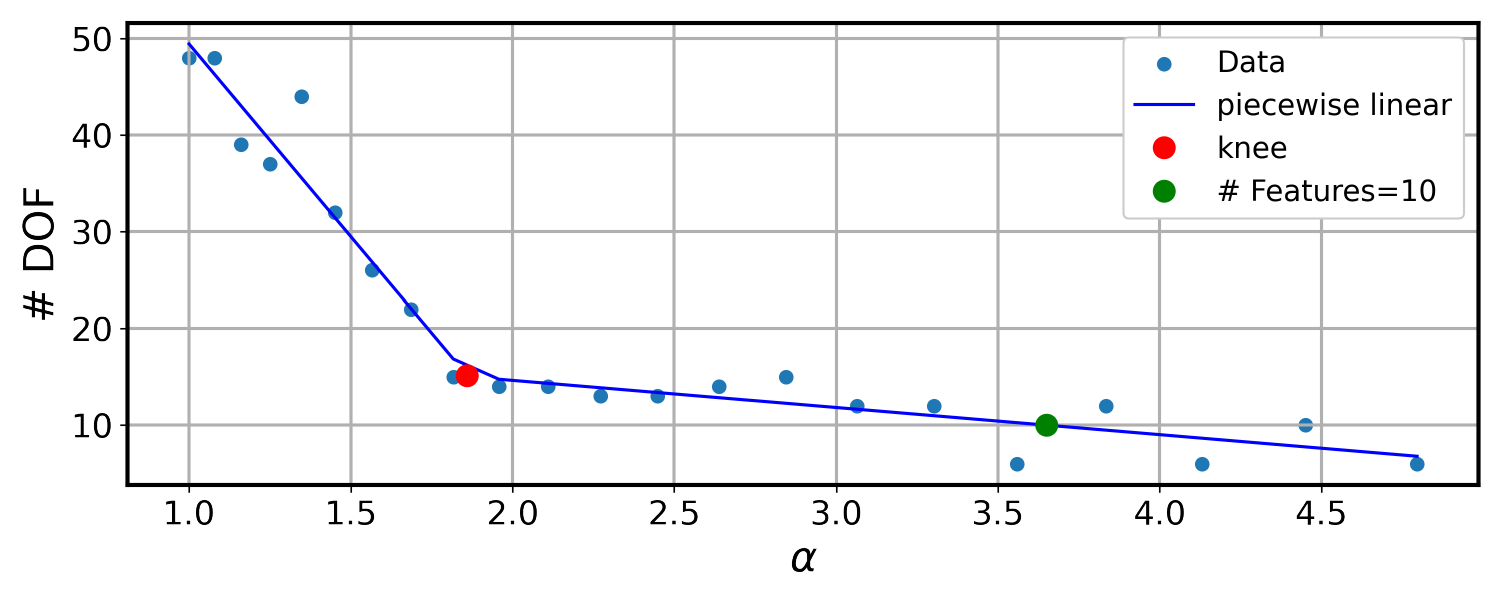}
  \caption{ The data shown in Figure \ref{fig_BNN_alpha_loop}b, fit by
    a piece-wise (bilinear) function. The approximate position of the
    ``knee'' in the curve is shown with the red symbol and the
    position on the curve ($\alpha=3.65$) where the number of degrees
    of freedom equals the number of features (\#DOF $=10$) is shown by
    the green symbol.  } \label{fig_BNN_alpha_loop_fits}
\end{figure}

It is illustrative to track the characteristics of our network as we
optimize it over a range of regularization parameters. In Figure
\ref{fig_BNN_alpha_loop} we show typical results for the data loss,
the training and testing accuracies, the AIC and BIC metrics, and the
estimated number of degrees of freedom. We typically see the
juxtaposition of two extremes, a dominance of the KL divergence term
for low $\alpha$ characterized by a rapid change with $\alpha$
transitioning through a relatively distinct ``knee'' to dominance by
the fit to data term characterized by a slow change. The ``knee'' in
the curves is distinct and indicates some balance between the extremes
yet is variable from one metric to the next and to our minds has no
obvious intrinsic significance. We note, however, that the number of
degrees of freedom curve provides us with a way to choose a value of
$\alpha$ both conservative and understandable. Recall that the
simplest decision boundary in our 10-feature problem would be a
hyperplane with 9 free parameters and that the VC dimension for a
linear perceptron in this case would be $d_{VC}=11$. We choose as a
compromise 10 ``effective'' degrees of freedom by our measure to give
a conservative solution to the problem, erring on the side of
underfitting the data. Figure \ref{fig_BNN_alpha_loop_fits} shows how
such a determination can be made. In our case we use the convention of
normalizing the KL divergence as the average per number of weights in
the network and normalizing the data fit as the data loss per training
example. With this convention the optimal value of regularization is
found to be $\alpha_{10} = 3.65$. We note that for a range of $\alpha$
around this number, none of the model behavior characteristics varies
greatly, and in fact the resulting predictive resource maps for this
range are all visually similar.

\subsubsection{Variability in Predictions}

\begin{figure}[p]
  \centering
  \includegraphics[width=0.8\textwidth]{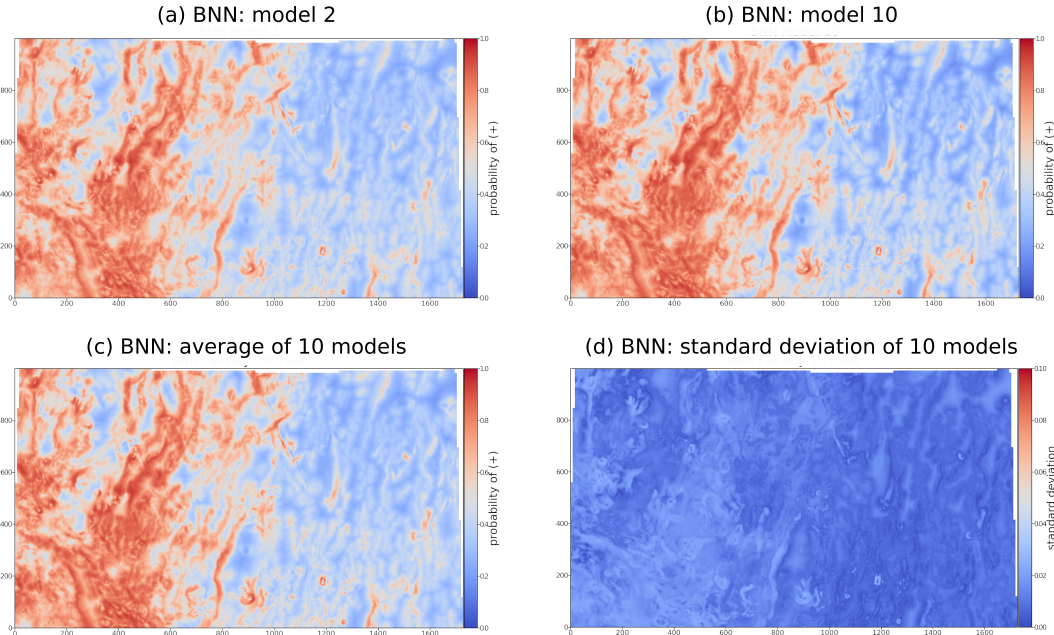}
  \caption{BNN model variability exposed by multiple runs. Each image
    represents an average of 128 samples from the posterior
    distribution of probabilities predicted by the neural network
    (which is an estimate of $\Pbar(\bfx)$, the mean of
    $f(p\cnd\bfx)$). BNN models \#2 (a) and \#10 (b) represent the
    extreme variability seen among 10 training trials. A comparison of
    the two extremes to the mean (c) and the standard deviation (d) of
    all 10 trials shows that there is very small variability both in
    absolute value and spatially in the Bayesian neural network
    compared to the deterministic case shown in Figure
    \ref{fig_ANN_variability}.}  \label{fig_BNN_variability}
\end{figure}

Given we now have a favored degree of regularization, we can explore
how well treating the problem as Bayesian has addressed one of the
issues seen during the application of deterministic neural networks to
this problem. Recall that the deterministic networks were found to
hold a noticeable degree of variability from one trial to the next,
which could not be easily understood or rectified. By including
uncertainty into the problem with the Bayesian approach we had hoped
to address this in a satisfactory way. Figure
\ref{fig_BNN_variability} shows the results. In the Bayesian neural
network case, variability among trials arises from random seeding of
the weights and biases distribution parameters $\mu$ and $\rho$ as
well as the need to randomly draw multiple samples from the posteriors
during prediction.  In comparing this directly to the variability from
the deterministic network trials (Figure \ref{fig_ANN_variability}), we
see that now there is very small variability both in absolute value
and spatially in the Bayesian neural network compared to the
deterministic case. We consider, then, that the variability seen
before has now been entrained into the way a Bayesian neural network
can quantify the spread or distribution in outcomes of the model. This
intrinsic variability now provides a measure of confidence in our
predictions as we will discuss in a following section.

\subsubsection{Synopsis}

In summary, we have implemented a variational Bayesian approach for
predicting geothermal energy resource potential in a stochastic
way. The interpretation of the predictions at each site is shown to be a
probability and conditioned on the evidence provided by the data. The
predictions have been made intentionally in a conservative fashion. By
optimizing the value of the regularization parameter we seek a model
that is only as complex (and confident in its predictions) as the
training data allow.

\subsection{Results of the Favored Model}

By using the Bayesian formulation for our problem, we have a final
optimized model, which is probabilistic in nature. This provides a
distinct advantage over a fixed-parameter neural network: that of
providing a range of predictions for each case we wish to test whose
spread represents a degree of reliability or confidence. Here we
illustrate this outcome explicitly and suggest how it might be used
to aid decision makers in the process of geothermal energy
resource assessment or prospecting.

\subsubsection{Distribution of Probabilities}

\begin{figure}[p]
  \centering
  \includegraphics[width=0.9\textwidth]{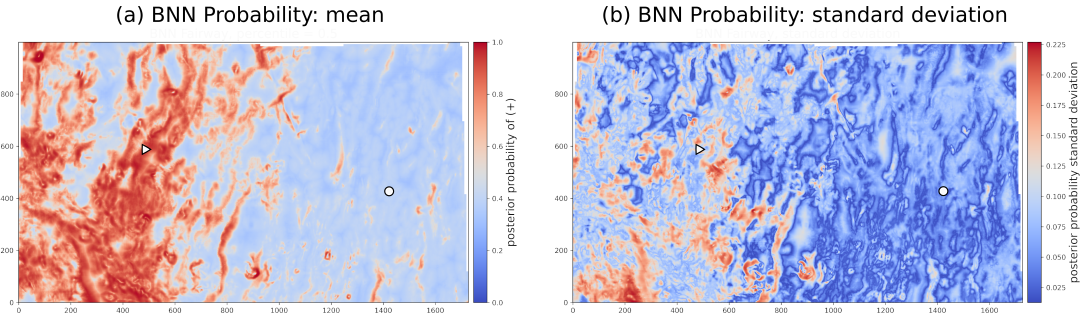}
  \caption{Geothermal resource potential model produced from a single
    training run at $\alpha_{10}=3.65$, the optimal value determined
    by consideration of model complexity described in Section
    \ref{sec:optimal}. Predictions used to make the maps represent the
    average of 8192 samples from the posterior distribution of
    probabilities predicted by the neural network (which is an
    estimate of $\Pbar(\bfx)$, the mean of $f(p\cnd\bfx)$). (a)
    represents the mean posterior probability and (b) represents the
    standard deviation of posterior probability. The white triangle
    and circle symbols mark the locations of the example positive and
    negative training sites, respectively, shown in Figure
    \ref{fig_BNN_example_pixel_posteriors}.}
    \label{fig_BNN_best_alpha}
\end{figure}

Our trained and optimized model was used to make predictions of the
probability of finding a viable geothermal resource (positive site)
within each pixel (geographic area of 250m by 250m square) for the
study area shown in Figure \ref{fig_PFA_study_area}. Two maps help to
illustrate the probabilistic nature of the result. Figure
\ref{fig_BNN_best_alpha}a shows the map for the mean of the
distribution at each pixel and Figure \ref{fig_BNN_best_alpha}b shows
the standard deviation. These maps together give a sense of the
geographic distribution of high and low geothermal resource
probabilities as well as a measure of the corresponding spread or
``uncertainty'' in those values.

\begin{figure}[p]
  \centering
  \includegraphics[width=0.9\textwidth]{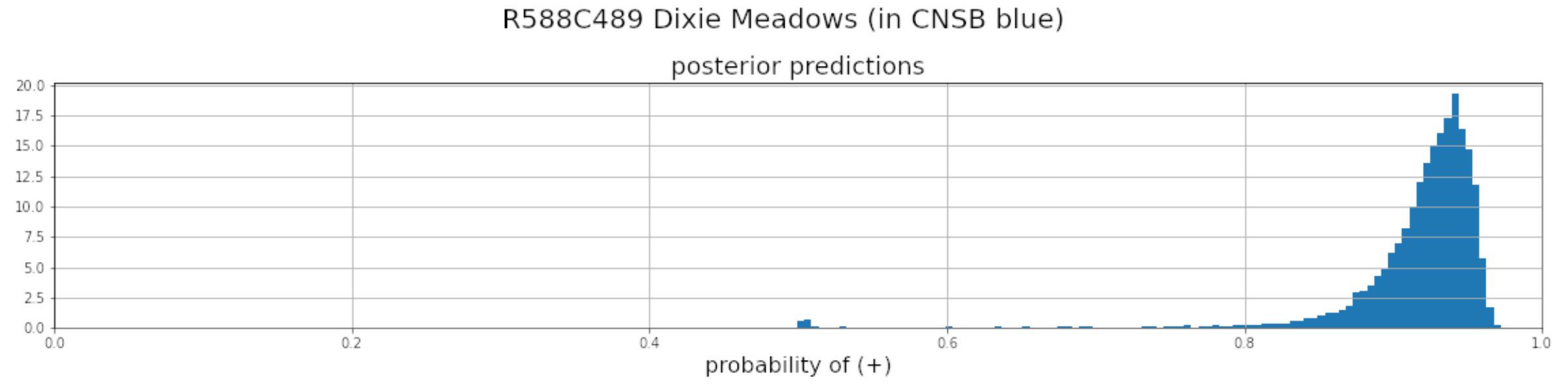}
  \includegraphics[width=0.9\textwidth]{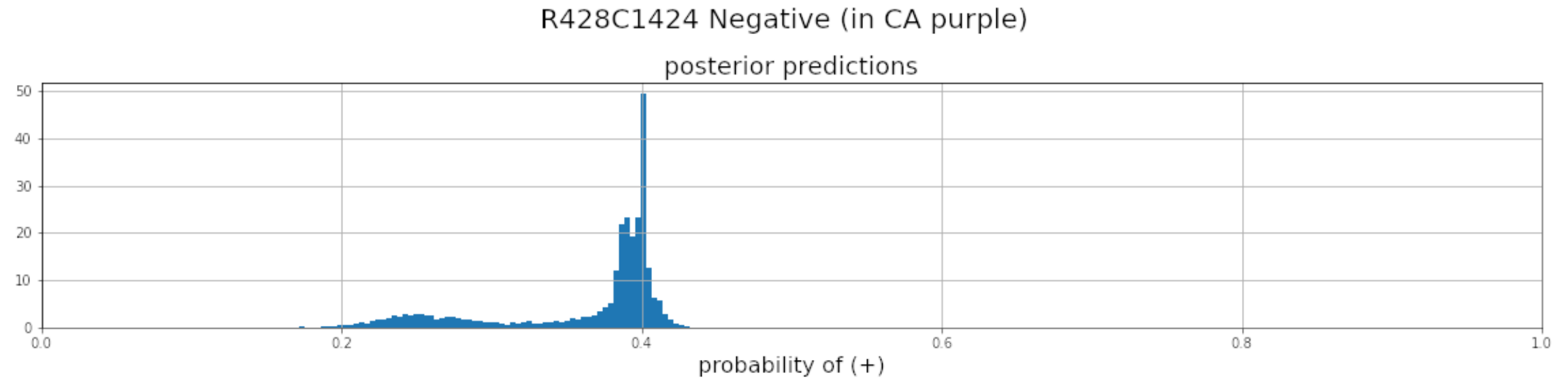} \\
  \caption{Examples of detailed posterior distributions $f(p\cnd\bfx)$ for
    two training sites. Top: The positive training site marked as a
    triangle in Figure \ref{fig_BNN_best_alpha}. Bottom: The negative
    training site marked as a circle in Figure
    \ref{fig_BNN_best_alpha}.}
  \label{fig_BNN_example_pixel_posteriors}
\end{figure}

These maps still do not tell the whole story. If one probes the
resource probability distribution at specific pixels, one by one, then
the true nature of this model are exposed. Figure
\ref{fig_BNN_example_pixel_posteriors} shows two typical distributions
of the resource probability for specific example pixels taken from the
training set: one positive site and one negative site. One can see the
distributions can be highly skewed and may be multi-modal. Therefore,
the simple mean and standard deviation maps are insufficient to
describe the model properties in detail. It is clear that one should
consider both the mean (which represents the average result if many
drilling trials could be made) and the spread or range of this
distribution to determine the variability that we expect to
encounter. This has important practical applications, especially in
cases where funding may allow only one chance to drill an exploratory
hole.

\subsubsection{A Tool for Decision Makers}

\begin{figure}[p]
  \centering
  \includegraphics[width=0.9\textwidth]{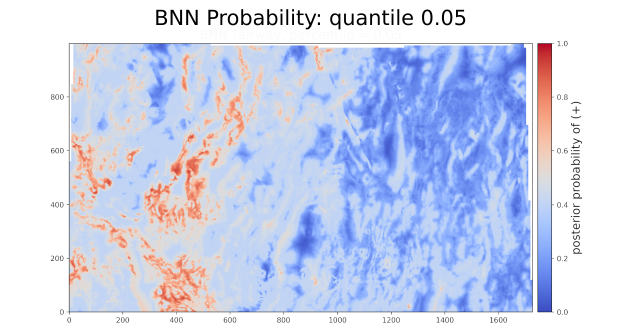}
  \caption{Posterior probability map for the $0.05$ quantile of
    $f(p\cnd\bfx)$ indicating where on the map 95\% of the models have
    at least the probability shown by the
    color.} \label{fig_BNN_best_alpha_0.05_percentile}
\end{figure}

Given that probing the Earth for geothermal energy is expensive, often
exceeding \$2M for drilling a geothermal production or injection well
\citep[e.g.][]{Lowry:2017}, we need to maximize our chances of
success. A practical way of looking at the model in this regard is as
follows. Looking at the individual distributions at specific pixels
such as in Figure \ref{fig_BNN_example_pixel_posteriors}, we realize
that our model actually represents a large family of models. If we
construct the cumulative distribution function at each site, a
specific quantile (or percentile) represents the number of the models
that agree on a specific outcome. For example, the resource
probability, say $p$, at the 0.5 quantile (50th percentile) represents
where half of the models have a resource probability $> p$ and half
have resource probability $< p$. More useful, for example, would be
the 0.05 quantile or the 5th percentile. There, 95\% of all models
have resource probability $> p$ and only 5\% smaller. A map
constructed in this way (Figure
\ref{fig_BNN_best_alpha_0.05_percentile}) is much more conservative
than the map of mean resource probabilities. High resource probability
locations shown on this 0.05 quantile map carry much higher confidence
and may be a practical tool for decision makers in the course of
prospecting for new resources.

\section{Summary}

We have considered the application of machine learning to the
evaluation of geothermal resource potential. A supervised learning
problem was posed in terms of a set of maps based on 10 geological and
geophysical features. These features were extracted from an extensive
study in Nevada and were then used to classify geographic regions
representing either a potential resource (+) or not (-). A training
set of positive (known resources or active power plants) and negative
training sites (known drill sites with unsuitable geothermal
conditions) was used to constrain and optimize various artificial
neural network models for this classification task. Among the models used,
we found that by using a Bayesian approach we could naturally include
model uncertainty and provide a way not only to predict resource
potential but to also give measures of confidence or reliability.

\paragraph{Acknowledgements.}
This work was supported in part by the U.S. Department of Energy’s
Office of Energy Efficiency and Renewable Energy (EERE) under the
Geothermal Technologies Program and Machine Learning Initiative award
number DE-EE0008762.

%\newpage

\bibliography{references}
\addcontentsline{toc}{section}{References}

\end{document}